\documentclass[a4paper,11pt]{article}
\usepackage{mathtools}
\usepackage{amsmath}
\usepackage{mathrsfs}
\usepackage{amssymb}
\usepackage{amsthm}
\usepackage{dsfont}
\usepackage{graphicx}
\usepackage{bmpsize}
\usepackage{times}
\usepackage{ textcomp }
\usepackage{mathtools}
\usepackage{latexsym, empheq, fancybox}
\usepackage{mathrsfs}
\usepackage{exscale}
\usepackage{booktabs, array}
\usepackage[english]{babel}
\usepackage{authblk}
\title{\Large\textbf{  Efficiency and Sensitivity Analysis of Observation Networks for Atmospheric Inverse Modelling with Emissions}
\thanks{This work was
supported by HITEC and IEK-8 of Forschungszentrum J\"ulich.}}

\author{Xueran Wu\thanks{IEK-8, Forschungszentrum J\"ulich, Wilhelm-Johnen-Stra\ss e,
52428, J\"ulich, Germany.
       {\tt\small Corresponding to xueranwu@uni-wuppertal.de}}$^{\ ,\ \mathsection}$,\    
Hendrik Elbern$^{\dagger,\ }$\thanks{Rhenish Institute for Environmental Research,  University of Cologne, Aachener Stra\ss e 209,
50931, Cologne, Germany. {\tt\small\quad he@eurad.uni-koeln.de}.} 
\ \ and Birgit Jacob\thanks{Mathematics Department, Unversity of Wuppertal,
Gau\ss stra\ss e 20, 42119, Wuppertal, Germany.  {\tt\small xueranwu@uni-wuppertal.de, jacob@math.uni-wuppertal.de}
        }
}

\begin{document}
\date{}

\maketitle
\begin{abstract}
The controllability of advection-diffusion systems, subject to uncertain initial values and emission rates, is estimated,
given sparse and error affected observations of prognostic state variables. In predictive geophysical model systems,
like atmospheric chemistry simulations,  different parameter families influence the temporal evolution of the system.
This renders initial-value-only optimisation by  traditional data assimilation methods as insufficient.
In this paper, a quantitative assessment method on validation of measurement configurations
to optimize initial values and emission rates,  and how to balance them, is introduced.
In this theoretical approach,  Kalman filter
and smoother and their ensemble based versions are combined with a singular
value decomposition, to evaluate the potential
improvement associated with specific observational network configurations. Further, with the same
singular vector analysis for the efficiency of observations, their
sensitivity to model control can be identified by determining the
direction and strength of maximum perturbation in a finite-time interval.
\end{abstract}

\section{Introduction}

Air quality and climate change are influenced by the fluxes of green house gases,
reactive gas emissions and aerosols in the atmosphere. The ability to quantify variable,
yet hardly observable emission rates is a key problem to be solved for the analysis of atmospheric systems,
and typically addressed by elaborate and costly field campaigns or permanently operational observation networks.
The temporal evolution of chemistry in the atmosphere is usually modelled by atmospheric chemistry transport models. Optimal simulations are based on techniques of combining numerical models with observations.
In meteorological forecast models, where initial values are insufficiently well known, while exerting a high influence on the model evolution, this procedure is termed data assimilation (\cite{Daley91}).
There is no doubt that the optimization of the initial state is always of great importance for the improvement of predictive skill.
However, especially for chemistry transport or greenhouse gas models with high dependence on the emissions in the troposphere, the optimization of initial state is no longer the only issue.
The lack of ability to observe and estimate surface emission fluxes directly with necessary accuracy
is a major roadblock, hampering the progress in predictive skills of climate and
atmospheric chemistry models. In order to obtain the Best Linear Unbiased Estimation (BLUE) from the model with observations,  efforts of optimization
included the emission rates by spatio-temporal data assimilation have been made. The first full chemical
implementation of the 4D-variational method for atmospheric chemistry initial values is introduced in \cite{Elbern99}. Further,  Elbern et al. (\cite{Elbern07}) took the strong constraint of the diurnal profile shape of emission rates such that their amplitudes and initial values
are the only uncertainty
to be optimized and then implemented it by 4D-variational inversion. This strong constraint approach is reasonable because the diurnal evolution of emissions are typically much better known than the
absolute amount of daily emissions. Moreover, several data assimilation strategies were designed to
adjust ozone initial conditions and emission rates separately or jointly in \cite{Tang11}. Bocquet et al. introduced a straightforward
extension of the iterative ensemble Kalman smoother in \cite{Bocquet13}.

In many cases, the better estimations of both the initial state and emission rates are not always
sustained based on appropriate observational network configurations when using popular data assimilation methods, such as 4D-variation and Kalman filter and smoother. It may hamper the optimization by unbalanced weights between the initial state and emission rates, which can, in practice, even result in degraded simulations beyond the time intervall with available observations. The ability to evaluate the suitability of an observational network
 to control chemical states and emission rates for its optimised designis the a key qualification, which needs to be adressed.

Singular value decomposition (SVD) can help identifying the priorities of observations by detecting
the fastest growing uncertainties. The targeted observations problem is an important topic in the field of
numerical weather prediction. Singular vector analysis based on SVD was firstly
introduced to numerical weather prediction by Lorenz (\cite{Lorenz65}), who applied
it to analyse the largest error growth rates in an idealised
atmospheric model. Because of the high cost of computation, the singular
vector analysis was not widely applied until 1980s. Later the method of
singular vector analysis of states of the meteorological model with high
dimension was feasible (\cite{Buizza95}).

In atmospheric chemistry, studies about the importance of observations are still sparse.
Khattatov et al. (\cite{Khattatov99}) firstly analysed the uncertainty of a
chemical compositions. Liao et al. (\cite{Liao06}) focused on the optimal placement
of observation locations of the chemical transport model. However,
singular vector analysis for atmospheric chemistry with emissions is
different since emissions play an similarly important role in forecast
accuracy with initial values.
Goris and Elbern (\cite{Goris13}) recently used
the singular vector decomposition to determine the sensitivity
of the chemical composition to emissions and initial values for a variety of chemical scenarios and integration length.

Hence, in this paper, applying the Kalman filter and smoother as the desirable data assimilation method we
introduce an approach to identify the sensitivities of a network to optimize emission rates and initial values independently and balanced prior to
any data assimilation procedure.
Through singular value decomposition and ensemble Kalman filter and smoother, the computational cost of this approach can be reduced so that it is feasible in practice.
Then, by the equivalence between 4D variation and Kalman filter for linear models, the approach is also feasible for the data assimilation of adjoint models via 4D variational techniques.

This paper is organized as follows.
In section \ref{model description}, we describe the atmospheric transport-diffusion model with emission rates first and then reconstruct the state
vector such that the emission rates are included dynamically.
In section \ref{efficiency}, the theoretical approach derives in order to determine the efficiency of observations or observational network configurations
before running any data assimilation procedure.
In section \ref{ensemble efficiency}, based on the theoretical analysis in section \ref{efficiency}, we discuss the ensemble approach to evaluate the efficiency
of observation configurations and present elementary examples.
In section \ref{sensitivity}, we present the approach to identify the sensitivity of
observations by determining the directions of maximum perturbation
growth to the initial perturbation. In the appendices, the above approaches are generalised to continuous-time systems for comprehensive applications.

\section{Model description}\label{model description}
\allowdisplaybreaks
The chemical tendency equation including emission rates, propagating forward in time, is usually described by the following atmospheric transport model
$$
\dfrac{dc}{dt}=\mathcal{A}(c)+e(t),
$$
where $\mathcal{A}$ is a nonlinear model operator, $c(t)$ and $e(t)$ are the state vector of chemical constituents and emission rates at time $t$, respectively .

The a priori estimate of the state vector of concentrations $c(t)$ is given and denoted by $c_b(t)$, termed background  state.
The a priori estimate of emission rates are usually taken from emission inventories, denoted by $e_b(t)$.

Let $\mathbf{A}$ be the tangent linear operator of $\mathcal{A}$, the evolution of the perturbation of states $c(t)$ and $e(t)$ follows the tangent linear model with $\mathbf{A}$ as
\begin{equation}\label{model}
\frac{d\delta c}{dt}=\mathbf{A}\delta c+\delta e(t),
\end{equation}
where $\delta c(t)$ is the perturbation evolving from the perturbation of initial state of chemical state $\delta c(t_0)=c(t_0)-c_{b}(t_0)$ and emission rates $\delta e(t)=e(t)-e_{b}(t)$.

After discretizing the tangent linear model in space, let $M(\cdot,\cdot)$ be the evolution operator or resolvent generated by $\mathbf{A}$. It is straightforward to obtain the linear solution of \eqref{model} with continuous time as
\begin{equation}\label{solution of model}
\delta c(t)=M(t,t_0)\delta c(t_0)+\int_{t_0}^t M(t, s)\delta e(s)ds,
\end{equation}
where $\delta c(t)\in \mathds R^{n}$, $\delta e(t)\in \mathds R^{n}$, $n$ the dimension of the partial phase space of concentrations and emission rates. Obviously, $M(\cdot,\cdot)\in \mathds R^{n\times n}$.

In addition, let $y(t)$ be the observation configuration of $c(t)$ and define
$$\delta y(t)=y(t)-\mathcal{H}(t) c_b(t),$$
where $\mathcal{H}(t)$ is a nonlinear forward observation operator mapping the model space to the observation space.
Then by linearising the nonlinear operator $\mathcal{H}$ as $H$, the linearised model equivalents of observation configurations can be presented as
\begin{equation*}
\delta y(t)=H(t)\delta c(t)+\nu(t),
\end{equation*}
where $\delta y(t)\in\mathds R^{m(t)}$, $m(t)$ the dimension of the phase space of observation configurations at time $t$.
$\nu(t)$ is the observation error at time $t$ of the Gaussian distribution with zero mean and variance $R(t)\in \mathds R^{m(t)\times m(t)}$.

It is feasible to apply the Kalman
filter and smoother into the model without any extension if the emission rates are accurate, which implies the initial state of concentration is the only parameter to be optimized. However, if the emission rates are poorly known, they should be combined into the state vector so that both of them can be updated
by a smoother application. To establish the model with a new combination of the initial state and emissions, let us rewrite the background of emission rates into the dynamic form
$$e_{b}(t)=M_e(t,s)e_{b}(s),$$
where $e_b(\cdot)$ is a $n$-dimensional vector of which the $i^{th}$ element is denoted by $e_b^i(\cdot)$ and
$M_e(t,s)$ is the diagonal matrix defined as
\begin{equation*}
M_e(t,s)=\left(\begin{array}{cccc}
      \frac{e_b^1(t)}{e_b^1(s)}& & &\\
        & \frac{e_b^2(t)}{e_b^2(s)} & & \\
      & & \ddots &\\
     & &  & \frac{e_b^n(t)}{e_b^n(s)}\\
      \end{array}
\right).
\end{equation*}

Since emission rates follow the diurnal variation, by taking the diurnal profile of emission rates as a constraint, the amplitude of emission rates can be estimated by constant emission factors (\cite{Elbern07}).
We reconstruct the dynamic model of emission rate perturbation as
\begin{equation}\label{emission}
\delta e(t)=M_e(t,s)\delta e(s).
\end{equation}
Then, \eqref{solution of model} can be written as
\begin{equation}\label{state}
 \delta c(t)=M(t,t_0)\delta c(t_0)+\int_{t_0}^t M(t, s)M_e(s,t_0)\delta e(t_0)ds.
\end{equation}
Hence, we obtain the extended model with emission rates
\begin{equation}\label{extended model}
\left(\begin{array}{c}
      \delta c(t)\\
     \delta e(t)
      \end{array}
\right)=
\left(\begin{array}{cc}
      M(t,t_0) & \int_{t_0}^t M(t, s)M_e(s,t_0)ds\\
      0     & M_e(t,t_0)
      \end{array}
\right)
\left(\begin{array}{c}
       \delta c(t_0)\\
     \delta e(t_0)
      \end{array}
\right).
\end{equation}

Typically, there is no direct observation for emissions. Therefore, we reconstruct the observation mapping as
\begin{equation*}
\delta y(t)=[H(t), 0_{n\times n}]\left(\begin{array}{c}
      \delta c(t)\\
     \delta e(t)
      \end{array}
\right)+\nu(t),
\end{equation*}
where $0_{n\times n}$ is a $n\times n$ matrix with zero elements.

It is clear now that both concentrations and emission rates are included into the state vector of the block extended model \eqref{extended model}, such that the Kalman smoother in a fixed time interval $[t_0,t_N]$ can be applied to optimize both of them.
 In our approach \eqref{emission}, the dynamic model of emission rates is forced to follow the background evolution of emission rates.
In fact, it was stated in several studies (\cite{Catlin09}, \cite{Gelb74}) that the best linear unbiased estimation (BLUE) of a random variable $x$, which implies this estimation, can minimise its variance.
The estimate via fix-interval Kalman smoother is the BLUE depending on all observation configurations in time interval $[t_0,t_N]$. In our case of emission rates,
the estimation of $e(t)$ by Kalman smoother on $[t_0,t_N]$ can be represented generally as the conditional expectation $E[e(t)\vert \{y(t), t\in [t_0,t_N]\}]$. By the linear property of conditional expectation,
\begin{eqnarray*}
&&E[e(t)\vert \{y(t), t\in [t_0,t_N]\}]\\&=&E[M_e(t,s)e(s)\vert \{y(t),t\in [t_0,t_N]\}]=M_e(t,s)E[e(s)\vert \{y(t), t\in [t_0,t_N]\}],
\end{eqnarray*}
which implies the dynamic model of emission rates \eqref{emission} satisfies the constraint of the diurnal shape of emission rates if $[t_0,t_N]$ covers 24 hours.

\section{Efficiency of observation networks of atmospheric inverse modelling with emission rates}\label{efficiency}
As mentioned before, the observational network configurations cannot necessarily help improving the initial state and emission rates in a balanced way.
If the estimation of both initial state and emission rates can be improved significantly,  we call the corresponding observation configurations as efficient or of high efficiency for both.
Otherwise, the observation configurations are only efficient to initial state or emission rates. However, it is usually difficult to foresee the efficiency of observation configurations.
Hence, the lack of the knowledge of the efficiency of observations may lead us to give the poor initial guesses and waste computational resource.
In this section, we will introduce the theoretical approach to determine the efficiency of observations via the Kalman filter and smoother in a finite-time interval.

\subsection{Theoretical analysis for the general discrete-time system}
For the application in atmospheric chemistry, let us consider the discrete-time system first.

Generalizing the extended atmospheric transport model with emission rates in a discrete time internal $[t_0,t_1,\cdots, t_N]$ to the following abstract linear system:
\begin{align}\label{discrete1}
&x(t_{k+1})=M(t_{k+1},t_k)x(t_k)+\varepsilon(t_k),\\
&y(t_k)=H(t_k)x(t_k)+\nu(t_k),
\end{align}
where $x(\cdot)\in \mathds R^n$ is the state variable, $y(t_k)\in \mathds R^{m(t_k)}$ is the observation vector at time $t_k$, the model error $\varepsilon(t_k)$ and the observation error
$\nu(t_k)$, $ k=1,\cdots, N$ follow  Gaussian distribution with zero mean and
$Q(t_k)$ and $R(t_k)$ are their covariance matrices respectively.

Denote the estimation of $x(t_k)$ based on $\{y(t_0),\cdots, y(t_k)\}$ by $\hat x(t_k\vert t_{k})$, termed as the analysis estimation, the estimation of $x(t_k)$ based on $\{y(t_0),\cdots, y(t_{k-1})\}$
by $\hat x(t_k\vert t_{k-1})$, termed as forecast estimation.
Correspondingly, $P(t_k\vert t_{k})$ and $P(t_k\vert t_{k-1})$ are the analysis error and forecast error covariance matrices of $\hat x(t_k\vert t_{k})$ and $\hat x(t_k\vert t_{k-1})$ respectively.
For convenience, the main results of the discrete-time Kalman filter can be summarised as follows:\\
(1) Analysis step:
\begin{equation*}
K(t_k)=P(t_k\vert t_{k-1})H^T(t_k)(H(t_k)P(t_k\vert t_{k-1})H^T(t_k)+R(t_k))^{-1};
\end{equation*}
\begin{equation*}
\hat x(t_k\vert t_{k})=\hat x(t_k\vert t_{k-1})+K(t_k)(y(t_k)-H(t_k)\hat x(t_k\vert t_{k-1}));
\end{equation*}
\begin{equation*}
P(t_k\vert t_{k})=(I-K(t_k)H(t_k))P(t_k\vert t_{k-1});
\end{equation*}
(2) Forecasting step:
\begin{equation*}
\hat x(t_{k+1}\vert t_{k})=M(t_{k+1},t_k)\hat x(t_k\vert t_{k});
\end{equation*}
\begin{equation*}
P(t_{k+1}\vert t_{k})=M(t_{k+1},t_k)P(t_k\vert t_{k})M^T(t_{k+1},t_k)+Q(t_k),
\end{equation*}
where for any matrix $M$, $M^T$ is the adjoint of $M$ and $M^{-1}$ is the inverse of $M$.

Denote the first guess of initial variance as $P(t_0\vert t_{-1})$ and select $P(t_0\vert t_{-1})$ and $R(t_k)$
to be symmetric and positive definite. Then we can rewrite
\begin{eqnarray*}
P(t_k\vert t_{k})&=&P(t_k\vert t_{k-1})\\&-&P(t_k\vert t_{k-1})H^T(t_k)(H(t_k)P(t_k\vert t_{k-1})H^T(t_k)+R(t_k))^{-1}H(t_k)P(t_k\vert t_{k-1}),
\end{eqnarray*}
and by the matrix inverse lemma (\cite{Woodbury50}), we have
\begin{equation}\label{inverse1}
P^{-1}(t_k\vert t_{k})=P^{-1}(t_k\vert t_{k-1})+H(t_k)^{T}R^{-1}(t_k)H(t_k).
\end{equation}
Further, assume the model error, which is usually unknown,  is negligible. Then,  we obtain
\begin{equation}\label{inverse2}
P^{-1}(t_{k+1}\vert t_{k})=M^{-T}(t_{k+1},t_k)P(t_k\vert t_{k})^{-1}M^{-1}(t_{k+1},t_k).
\end{equation}
Hence, by the deduction based on \eqref{inverse1} and \eqref{inverse2}, we have
 \begin{eqnarray*}
&&P^{-1}(t_{k+1}\vert t_{k})\\
&=&M^{-T}(t_{k+1},t_k)P^{-1}(t_k\vert t_{k-1})M^{-1}(t_{k+1},t_k)\\
&&+M^{-T}(t_{k+1},t_k)H^{T}(t_k)R^{-1}(t_k)H(t_k)M^{-1}(t_{k+1},t_k)\\
&=&M^{-T}(t_{k+1},t_{k-1})P^{-1}(t_{k-1}\vert t_{k-2})M^{-1}(t_{k+1},t_{k-1})\\
&&+M^{-T}(t_{k+1},t_{k-1})H^{T}(t_{k-1})R^{-1}(t_{k-1})H(t_{k-1})M^{-1}(t_{k+1},t_{k-1})\\
&&+M^{-T}(t_{k+1},t_k)H^{T}(t_k)R^{-1}(t_k)H(t_k)M^{-1}(t_{k+1},t_k)\\
&=&M^{-T}(t_{k+1},t_0)P^{-1}(t_0\vert t_{-1})M^{-1}(t_{k+1},t_0)\\
&&+\sum_{i=0}^{k}M^{-T}(t_{k+1},t_i)H^{T}(t_i)R^{-1}(t_i)H(t_i)M^{-1}(t_{k+1},t_i).
  \end{eqnarray*}

Define $\hat x(t_0\vert t_{k})=E[x(t_0)\vert y(t_0),\dots, y(t_k)]$ and denote its covariance matrix as
$$P(t_0\vert t_k)=E[(x(t_0)-\hat x(t_0\vert t_{k}))(x(t_0)-\hat x(t_0\vert t_{k}))^T],$$
which, according to the definition of $\hat x(t_0\vert t_{k})$, is the covariance of the estimate of the state from the fixed-interval Kalman smoother.
Then
\begin{eqnarray}\label{inverse cov}
&&P^{-1}(t_0\vert t_{k})\\\nonumber
&=&E[M^{-1}(t_{k+1},t_0)(x(t_{k+1})-\hat x(t_{k+1}\vert t_{k}))(x(t_0)-\hat x(t_{k+1}\vert t_{k}))^T M^{-T}(t_{k+1},t_0)]^{-1}\\\nonumber
&=& M^{T}(t_{k+1},t_0)P^{-1}(t_{k+1}\vert t_{k})M(t_{k+1},t_0)\\\nonumber
&=& P^{-1}(t_0\vert t_{-1})+\sum_{i=0}^{k}M^{T}(t_{i},t_0)H^{T}(t_i)R^{-1}(t_i)H(t_i)M(t_{i},t_0).
\end{eqnarray}

In particular, for $k=N$, taking the observations in the entire time interval into account, we have
\begin{equation}\label{inverse smoother}
P^{-1}(t_0\vert t_{N})= P^{-1}(t_0\vert t_{-1})+\sum_{i=0}^{N}M^{T}(t_{i},t_0)H^{T}(t_i)R^{-1}(t_i)H(t_i)M(t_{i},t_0).
\end{equation}

It is clear that \eqref{inverse smoother} includes all known information of the model with initial variance and observation configurations before any data assimilation procedure.
At the same time, it is independent of any specific data and states. Actually, if we define
\begin{equation}\label{def G}
\mathcal{G}=\left(\begin{array}{cc}
      H(t_0)M(t_0,t_0)\\
      H(t_1)M(t_1,t_0)\\
      \vdots\\
      H(t_N)M(t_N,t_0)
      \end{array}
\right)
\end{equation}
and
\begin{equation*}
\mathcal R^{-1}=\left(\begin{array}{ccccc}
       R^{-1}(t_0)& & &\\
       &  R^{-1}(t_1)& &\\
      & &\ddots&\\
      & & & R^{-1}(t_N)
      \end{array}
\right),
\end{equation*}
\eqref{inverse smoother} can be written as
\begin{equation}\label{inverse smoother2}
P^{-1}(t_0\vert t_N)= P^{-1}(t_0\vert t_{-1})+\mathcal{G}^T\mathcal R^{-1}\mathcal{G},
\end{equation}
where $\mathcal{G}^TR^{-1}\mathcal{G}$ equals the observability Gramian with $\mathcal R^{-1}$ from control theory, which is appropriate for how well states of a model can be inferred by the external observations.

Though \eqref{inverse smoother2} meets the demand to represent the covariance by all known information before starting the data assimilation procedure, the statistic interpretation of the inverse of covariance is still blurred to the application. Therefore,
for evaluating the improvement of the estimation with the initial variance $P(t_0\vert t_{-1})$, we define  \emph{relative improvement covariance} as
\begin{eqnarray}\label{improve normal}
&&P^{-\frac{1}{2}}(t_0\vert t_{-1})(P(t_0\vert t_{-1})-P(t_0\vert t_N))P^{-\frac{1}{2}}(t_0\vert t_{-1})\nonumber\\&=&I-P^{-\frac{1}{2}}(t_0\vert t_{-1})P(t_0\vert t_N)P^{-\frac{1}{2}}(t_0\vert t_{-1}),
\end{eqnarray}
where $I$ is the identity matrix.

The above improvement covariance is a normalised matrix of the difference between the initial variance $P(t_0\vert t_{-1})$ and the covariance matrix $P(t_0\vert t_N)$ from Kalman smoother. Especially,  $P^{-\frac{1}{2}}(t_0\vert t_{-1})P(t_0\vert t_N)P^{-\frac{1}{2}}(t_0\vert t_{-1})$ can be understood as the covariance matrix from the fixed-interval Kalman smoother normalised by the initial variance.
The symmetric normalised matrix guarantees the improvement covariance to be positive-definite. Further, its singular values or the eigenvalues are bounded since the sum of the eigenvalues of a matrix is equal to its trace.
In fact, from \eqref{improve normal}, we have
\begin{equation*}
0\leqslant P^{-\frac{1}{2}}(t_0\vert t_{-1})(P(t_0\vert t_{-1})-P(t_0\vert t_N))P^{-\frac{1}{2}}(t_0\vert t_{-1})<I,
\end{equation*}
which implies that its trace is always less than the dimension of the state vector of the model.

Since $P(t_0\vert t_N)$ is unknown before the data assimilation procedure is finished, we rewrite the relative improvement covariance as
\begin{eqnarray}\label{improve measure}
&&P^{-\frac{1}{2}}(t_0\vert t_{-1})(P(t_0\vert t_{-1})-P(t_0\vert t_N))P^{-\frac{1}{2}}(t_0\vert t_{-1})\nonumber\\
&=&P^{-\frac{1}{2}}(t_0\vert t_{-1})(P(t_0\vert t_{-1})-(P^{-1}(t_0\vert t_{-1})+\mathcal{G}^{T}\mathcal R^{-1}\mathcal{G})^{-1})P^{-\frac{1}{2}}(t_0\vert t_{-1})\nonumber\\
&=&I-P^{-\frac{1}{2}}(t_0\vert t_{-1})(P^{-1}(t_0\vert t_{-1})+\mathcal{G}^{T}\mathcal R^{-1}\mathcal{G})^{-1}P^{-\frac{1}{2}}(t_0\vert t_{-1})\nonumber\\
&=&I-(I+P^{\frac{1}{2}}(t_0\vert t_{-1})\mathcal{G}^{T}\mathcal R^{-1}\mathcal{G}P^{\frac{1}{2}}(t_0\vert t_{-1}))^{-1}.
\end{eqnarray}

It is clear to see from \eqref{improve measure} the improvement covariance defined in \eqref{improve normal} that 
$$I+P^{\frac{1}{2}}(t_0\vert t_{-1})\mathcal{G}^{T}\mathcal R^{-1}\mathcal{G}P^{\frac{1}{2}}(t_0\vert t_{-1})$$ is still invertible even without observability of the system,
which means $\mathcal{G}^{T}\mathcal{G}$ is not full-rank.
However, it is not easy to calculate \eqref{improve measure} directly. Hence, by singular value decomposition,
\begin{equation*}
P^{\frac{1}{2}}(t_0\vert t_{-1})\mathcal{G}^{T}\mathcal R^{-\frac{1}{2}}=VS U^T,
\end{equation*}
where $V$ and $U$ are  unitary matrices consisted of the left and right singular vectors, $S$ is the rectangular diagonal matrix consisting of the singular values.

We can simplify \eqref{improve measure} as
\begin{eqnarray}\label{improve measure2}
&&P^{-\frac{1}{2}}(t_0\vert t_{-1})(P(t_0\vert t_{-1})-P^{-\frac{1}{2}}(t_0\vert t_N))P(t_0\vert t_{-1})\nonumber\\
&=&I-(I+P^{\frac{1}{2}}(t_0\vert t_{-1})\mathcal{G}^{T}\mathcal R^{-1}\mathcal{G}P^{\frac{1}{2}}(t_0\vert t_{-1}))^{-1}\nonumber\\
&=&I-(I+VSS^T V^T)^{-1}\nonumber\\
&=& VV^T-(VV^T+VSS^T V^T)^{-1}\nonumber\\
&=& VV^T-(V(I+SS^T) V^T)^{-1}\nonumber\\
&=& V(I-(I+SS^T)^{-1}) V^T\nonumber\\
&=& \sum_{i=1}^{r}\dfrac{s_i^2}{1+s_i^2}v_{i}v_{i}^T,
\end{eqnarray}
where $r$ is the rank of \eqref{improve measure} and $v_{i}$ is the $i^{th}$ left singular vector in $V$ related to the singular value $s_i$, which is the $i^{th}$ element on the diagonal of $S$.

Let us consider the improvement of each element in the state vector as the corresponding value in the diagonal of the relative improvement covariance.
From \eqref{improve measure2}, we denote the relative improvement of $j^{th}$ element in $x(t_0)$ as $\tilde P_{j}$, then,
$$\tilde P_j=\sum_{i=1}^{r}\dfrac{s_i^2}{1+s_i^2}(v_{ij})^2,$$
where $v_{ij}$ is the $j^{th}$ element of $v_i$.

Get a deeper insight into the capacity of the observation networks to improve the estimation of all states of the model, some important indices need to be considered.
In fact, in order to evaluate the total improvement of the model, the nuclear norm for matrices, or equivalently, the 1-norm is appropriate, which is defined as
\begin{equation*}
\Vert M \Vert_{1}=\text{tr}(\sqrt{M^T M}),
\end{equation*}
where $M$ is any matrix and $\text{tr}(\cdot)$ denote the trace of the matrix.

For \eqref{improve measure2}, we denote
\begin{equation}\label{denote P}
\tilde P=P^{-\frac{1}{2}}(t_0\vert t_{-1})(P(t_0\vert t_{-1})-P(t_0\vert t_N))P^{-\frac{1}{2}}(t_0\vert t_{-1}),
\end{equation}
 according to \eqref{improve measure2},
\begin{equation*}
\Vert \tilde P\Vert_1= \sum_{i=1}^{r}\dfrac{s_i^2}{1+s_i^2},
\end{equation*}
which is called the \textit{total improvement value}.

As we mentioned before, $\Vert \tilde P\Vert_1<\Vert I\Vert_1=n$,
where $n$ can be considered as the total improvement value, if the system is fully known, which implies the optimal estimation is the value of state. Thus, if we consider the ratio
\begin{equation}\label{relative degree}
\tilde p=\dfrac{\Vert \tilde P\Vert_1}{\Vert I\Vert_1}=\dfrac{\Vert \tilde P\Vert_1}{n}\in[0,1),
\end{equation}
the percentage of the total improvement of the model is obtained, which is called the \textit{relative improvement degree}.

\subsection{Application to the extended atmospheric transport model with emissions}\label{extended model with emiss}
For the atmospheric transport model extended with emissions, composing the dimension of the original state $c\in \mathds R^n$ and emission rates $e\in \mathds R^n$ respectively,  we divide  \eqref{denote P} into a block matrix
\begin{equation*}
\tilde P
=\left(\begin{array}{cc}
      \tilde P^c & \tilde P^{ce} \\
      \tilde P^{ec} & \tilde P^e
      \end{array}\right)
= \sum_{i=1}^{2n}\dfrac{s_i^2}{1+s_i^2}\left(\begin{array}{c}
                                                      v_{i}^c\\
                                                      v_{i}^e
                                                     \end{array}\right)(v_{i}^{{c}^T}, v_{i}^{{e}^T})\in\mathds R^{2n\times 2n},
\end{equation*}
where $P^c$ is the relative improvement covariance of the state $c(t_0)$, $P^e$ is the relative improvement covariance of the emission rates $e(t_0)$,
$P^{ce}=(P^{ec})^T$ is the relative improvement covariance between $c(t_0)$ and $e(t_0)$ and $(v_i^{c^T},v_i^{e^T})^T=v_i$.

Then, it is easy to calculate
\begin{equation*}
\tilde P^c= \sum_{i=1}^{2n}\dfrac{s_i^2}{1+s_i^2} v_{i}^cv_{i}^{{c}^T},
\quad
\tilde P^e= \sum_{i=1}^{2n}\dfrac{s_i^2}{1+s_i^2} v_{i}^ev_{i}^{{e}^T}.
\end{equation*}

Further, the relative improvements of $j^{th}$ element in $c(t_0)$ and $e(t_0)$ are
 \begin{equation*}
\tilde P_j^c= \sum_{i=1}^{2n}\dfrac{s_i^2}{1+s_i^2} (v_{ij}^{c})^2,
\quad
\tilde P_j^e= \sum_{i=1}^{2n}\dfrac{s_i^2}{1+s_i^2} (v_{ij}^{e})^2,
\end{equation*}
where $v_{ij}^{c}$ and  $v_{ij}^{e}$ are respectively the $j^{th}$ element of $v_i^c$ and $v_i^e$.

Moreover, the total improvement values of concentration and emission rates are
\begin{equation*}
\Vert \tilde P^c \Vert_{1}= \sum_{i=1}^{2n}\dfrac{s_i^2}{1+s_i^2} \text{tr}(v_{i}^cv_{i}^{{c}^T}),
\quad
\Vert \tilde P^e \Vert_{1}= \sum_{i=1}^{2n}\dfrac{s_i^2}{1+s_i^2} \text{tr}(v_{i}^ev_{i}^{{e}^T}).
\end{equation*}

It is worth noting that $$\tilde P^c=(P^{c}(t_0\vert t_{-1}))^{-\frac{1}{2}}(P^{c}(t_0\vert t_{-1})-P^c(t_0\vert t_N))(P^{c}(t_0\vert t_{-1}))^{-\frac{1}{2}}$$ and
$$\tilde P^e=(P^{e}(t_0\vert t_{-1}))^{-\frac{1}{2}}(P^{e}(t_0\vert t_{-1})-P^e(t_0\vert t_N))(P^{e}(t_0\vert t_{-1}))^{-\frac{1}{2}}$$
if and only if there is no correlation between the initial concentration and emission rates.
In fact, if we assume $P^{ce}(t_0\vert t_{-1})=0_{n\times n}$, the corresponding relative improvement degrees of concentration and emission rates are defined as
\begin{equation*}
\tilde p^{c}=\frac{\Vert \tilde P^{c}\Vert_1}{n},
\quad
\tilde p^{e}=\frac{\Vert \tilde P^{e}\Vert_1}{n}.
\end{equation*}

According to \eqref{relative degree}, it is obvious that $\tilde p^{c}\in [0,1)$ and $\tilde p^{e}\in [0,1)$ show the percentages of the relative improvements of concentration and emission rates, respectively.
However, since
\begin{equation*}
\frac{\Vert \tilde P^c\Vert_1}{n}+\frac{\Vert \tilde P^e\Vert_1}{n}>1,
\end{equation*}
which indicates the normalisation is just with respect to the extended covariance matrix rather than specified to the state $c$ and emission rates $e$.
The relative improvement degree cannot serve our objective to distinguish the observability of concentration and emission rates and balance them quantitatively. However,
by observing the block form of $\tilde P$, it is easy to obtain
\begin{equation*}
\Vert \tilde P^c \Vert_{1}+\Vert \tilde P^e \Vert_{1}=\Vert \tilde P \Vert_{1}.
\end{equation*}
For comparing the improvement of the concentration and emission rates, we define \emph{relative improvement ratios} for the state or the emission rates as
\begin{equation*}
\tilde p^c=\frac{\Vert \tilde P^c\Vert_1}{\Vert \tilde P \Vert_{1}},
\quad
\tilde p^e=\frac{\Vert \tilde P^e\Vert_1}{\Vert \tilde P \Vert_{1}}, \quad
\tilde p^e+\tilde p^c\equiv 1.
\end{equation*}

In a sum, if the total improvement value or relative improvement degree of the model is almost zero,
the relative improvement ratios do not need to be considered since no state of the model is improved.
Otherwise, $\{\tilde P_j^c\}_{j=1}^{n}$ and  $\{\tilde P_j^e\}_{j=1}^{n}$, which show the improvement of each parameter $j$ of concentrations and emission rates respectively, can help us determining which parameters can be optimized by the existing observation configurations.
By comparing $\tilde p^c$ with $\tilde p^e$, it is clear that the estimation of the one with the larger ratio (larger magnitude of the improvement covariance) can be improved more efficiently by the existing observation configurations.
In other words, if $\tilde p^c>\tilde p^e$, the existing observation configurations are more sensible to the initial values of concentration. Conversely, if $\tilde p^c <\tilde p^e $, the observation configurations can help improving the estimation of emission rates more. According to $\tilde p^c$ and $\tilde p^e$, the 'weight' between the concentrations and emission rates can be decided quantitatively.
In a data assimilation context, where observations are in a weighted relation to the background, the BLUE favours the more sensitive parameters.

In a special case that $\tilde p^e$ is very close to zero, the emission rates can be viewed as the input in the model without any optimization when the data assimilation procedure is started.

For the completion of the theorem and wider application, the generalisation of the above method for the continuous-time system is introduced in Appendix A.
 While the derivation is totally different from the discrete-time system, it can be shown that the theoretical analysis still holds for the continuous-time system.

\section{Application to the ensemble Kalman filter and smoother}\label{ensemble efficiency}
In practice, the standard Kalman filter and smoother cannot be applied directly to transport modells due to their computational complexity.
The ensemble Kalman filter (EnKF) and smoother (EnKS), as a Monte Carlo implementation originating from Kalman filter and smoother, are suitable for problems with a large number of control variables
and are an important tool in the field of data assimilation (\cite{Evensen09}). In this section, we will introduce how to apply the above method if an Ensemble Kalman filer and smoother are applied as the data assimilation method.

For the abstract discrete-time system \eqref{discrete1}, we denote the ensemble samples of $\hat x(t_i\vert t_{i-1})$ and $\hat x(t_i\vert t_{i})$ , $i=1,\cdots,N$ respectively by
\begin{align*}
 X(t_i\vert t_{i-1})=(\hat x_1(t_i\vert t_{i-1}),\hat x_2(t_i\vert t_{i-1}),\cdots,\hat x_q(t_i\vert t_{i-1})),\\
 X(t_i\vert t_{i})=(\hat x_1(t_i\vert t_{i}),\hat x_2(t_i\vert t_{i}),\cdots,\hat x_q(t_i\vert t_{i})),
\end{align*}
where $q$ is the number of ensemble members.

Correspondingly, their ensemble means are
\begin{align*}
& \bar x(t_i\vert t_{i-1})=\frac{1}{q}\sum_{k=1}^q \hat x_{k}(t_i\vert t_{i-1})=\dfrac{1}{q}X(t_i\vert t_{i-1})\mathds 1_{q\times 1},\\
& \bar x(t_i\vert t_{i})=\frac{1}{q}\sum_{k=1}^q \hat x_{k}(t_i\vert t_{i})=\dfrac{1}{q}X(t_i\vert t_{i})\mathds 1_{q\times 1},
\end{align*}
where $\mathds 1_{i\times j}$ is a $i\times j$ matrix where each element is equal to $1$.

Note the ensemble perturbation matrix consist of the perturbation of each sampling by
\begin{equation*}
\tilde X(t_i\vert t_{i-1})= X(t_i\vert t_{i-1})-\dfrac{1}{q}X(t_i\vert t_{i-1})\mathds 1_{q\times q}.
\end{equation*}
Then the ensemble covariance is
\begin{equation}\label{ensemble cov}
 \bar P(t_i\vert t_{i-1})=\dfrac{1}{q-1}\tilde X(t_i\vert t_{i-1})\tilde X^T(t_i\vert t_{i-1}),
\quad
 \bar P(t_i\vert t_{i})=\dfrac{1}{q-1}\tilde X(t_i\vert t_{i})\tilde X^T(t_i\vert t_{i}).
\end{equation}

Further, we define the ensemble observation equivalent in the entire time interval in observation space as
\begin{equation*}
 y_k^{f}=\mathcal G\hat x_k(t_0\vert t_{-1}),\ \ k=1,\cdots,q
\end{equation*}
and denote the ensemble mean and the forecast error covariance matrix in observation space by
\begin{equation*}
\bar y^f=\frac{1}{q}\sum_{k=1}^q y_k^{f}, \quad
\bar P_{yy}^f=\frac{1}{q-1}\sum_{k=1}^q(\hat y_k^f-\bar y^f)(\hat y_k^f-\bar y^f)^T=\mathcal G\bar P(t_0\vert t_{-1})\mathcal G^T.
\end{equation*}
Similarly, we denote the ensemble covariance between the initial states and the forecasting observations by
 \begin{equation*}
\bar P_{xy}^f=\frac{1}{q-1}\sum_{k=1}^q(\hat x_k(t_0\vert t_{-1})-\bar x(t_0\vert t_{-1}))(\hat y_k^f-\bar y^f)^T=\bar P(t_0\vert t_{-1})\mathcal G^T.
\end{equation*}

In addition, we define the ensemble observations as
\begin{equation*}
\hat y_k(t_i)=y(t_i)+\nu_k(t_i), \ \ k=1,\cdots,q, \ \ i=1,\cdots,N
\end{equation*}
and assume
$
\bar \nu(t_i)=\frac{1}{q}\sum_{k=1}^q\nu_k(t_i)=0.
$
Denote the ensemble covariance of observation errors by
$
\bar R(t_i)=\frac{1}{q-1}\sum_{k=1}^q\nu_k(t_i)\nu_k^T(t_i).
$
Further, we assume
\begin{equation*}
\mathcal{\bar R}^{-1}=\left(\begin{array}{ccccc}
      \bar R^{-1}(t_0)& & &\\
       &  \bar R^{-1}(t_1)& &\\
      & &\ddots&\\
      & & &\bar R^{-1}(t_N)
      \end{array}
\right).
\end{equation*}

It is shown by Evensen (\cite{Evensen09}) that the ensemble forecasting and analysis covariances have the same form with the covariances in the standard Kalman filter.
It indicates that \eqref{inverse1} and \eqref{inverse2} are also true for $\bar P(t_i\vert t_{i})$ and $\bar P(t_i\vert t_{i-1})$. So upon substituting $\bar P(t_0\vert t_{-1})$ into
$P(t_0\vert t_{-1})$ in \eqref{inverse smoother2}, according to matrix inversion lemma, we obtain
\begin{eqnarray}\label{ensemble inverse cov}
&&\bar P(t_0\vert t_N)\nonumber\\&=&\bar P(t_0\vert t_{-1})-\bar P(t_0\vert t_{-1})\mathcal{G}^T\mathcal R^{-\frac{1}{2}}(I+\mathcal R^{-\frac{1}{2}}\mathcal G\bar P(t_0\vert t_{-1})\mathcal G^T\mathcal R^{-\frac{1}{2}})^{-1}\mathcal R^{-\frac{1}{2}}\mathcal{G}\bar P(t_0\vert t_{-1}\nonumber)\\
&=& \bar P(t_0\vert t_{-1})-\bar P_{xy}^f\mathcal R^{-\frac{1}{2}}(I+\mathcal R^{-\frac{1}{2}}\bar P_{yy}^f\mathcal R^{-\frac{1}{2}})^{-1}\mathcal R^{-\frac{1}{2}}(\bar P_{xy}^f)^T.
\end{eqnarray}
Then,
\begin{eqnarray}\label{ensemble evaluation}
 &&\bar P^{-\frac{1}{2}}(t_0\vert t_{-1})(\bar P(t_0\vert t_{-1})-\bar P(t_0\vert t_N))\bar P^{-\frac{1}{2}}(t_0\vert t_{-1})\nonumber\\
 &=&\bar P^{-\frac{1}{2}}(t_0\vert t_{-1})\bar P_{xy}^f\mathcal R^{-\frac{1}{2}}(I+\mathcal R^{-\frac{1}{2}}\bar P_{yy}^f\mathcal R^{-\frac{1}{2}})^{-1}\mathcal R^{-\frac{1}{2}}(\bar P_{xy}^f)^T\bar P^{-\frac{1}{2}}(t_0\vert t_{-1}).
\end{eqnarray}

To simplify \eqref{ensemble evaluation}, let $\sum_{i=1}^{N}m(t_i)=m$, by singular value decomposition, we have
\begin{equation}\label{ensemble svd}
\bar P^{-\frac{1}{2}}(t_0\vert t_{-1})\bar P_{xy}^f\mathcal R^{-\frac{1}{2}}=VSU^T\in \mathds R^{n\times m}
\end{equation}
where $U\in \mathds R^{m\times m}$ consists of the eigenvectors of $\mathcal R^{-\frac{1}{2}}\mathcal{G}\bar P(t_0\vert t_{-1})\mathcal{G}^T\mathcal R^{-\frac{1}{2}}$,
$V\in\mathds R^{n\times  n}$ consists of the eigenvectors of $\bar P^{\frac{1}{2}}(t_0\vert t_{-1})\mathcal{G}^T\mathcal R^{-1}\mathcal{G}\bar P^{\frac{1}{2}}(t_0\vert t_{-1})$,
$S\in \mathds R^{n\times m}$ consists of the singular values.

Let $r$ be the rank of \eqref{ensemble svd}, the \emph{ensemble relative improvement covariance} is defined as
\begin{eqnarray}\label{ensemble improve}
 &&\bar P^{-\frac{1}{2}}(t_0\vert t_{-1})(\bar P(t_0\vert t_{-1})-\bar P(t_0\vert t_N))\bar P^{-\frac{1}{2}}(t_0\vert t_{-1})\nonumber\\
 &=&VS^TU^T(UU^T+U(SS^T)U^T)^{-1}US V^T\nonumber\\
 &=&VS^T(I+S^TS)^{-1}S V^T\nonumber\\
 &=&\sum_{i=1}^{r}\dfrac{s_i^2}{1+s_i^2}v_{i}v_{i}^T
\end{eqnarray}
and its diagonal shows \emph{ensemble relative improvements} of states.

Obviously, \eqref{ensemble improve} has the same form as \eqref{improve measure2}. In fact, for the linear dynamic model, we have
\begin{equation}\label{ensemble equivalent}
\bar P^{-\frac{1}{2}}(t_0\vert t_{-1})\bar P_{xy}^f\mathcal{\bar R}^{-\frac{1}{2}}=\bar P^{\frac{1}{2}}(t_0\vert t_{-1})\mathcal G^T \mathcal{\bar R}^{-\frac{1}{2}},
\end{equation}
which implies if we just substitute $\mathcal R$ and $P(t_0\vert t_{-1})$ in \eqref{improve measure2} by $\mathcal{\bar R}$ and $\bar P(t_0\vert t_{-1})$,
the final results of \eqref{improve measure2} and \eqref{ensemble improve} are equivalent.
However, it is much more efficient to calculate practically the singular values and vectors of the left-side matrix of \eqref{ensemble equivalent} than to calculate the singular
values and vectors of the right-side matrix of \eqref{ensemble equivalent}, since the explicit calculation of $\mathcal G$ is not necessary.

Further on, it is worth noticing that if the ensemble size $q$ is less than the dimension of the model $n$,
the initial ensemble covariance  $\bar P(t_0\vert t_{-1})$ is not invertible. In this case, it is reasonable to replace $\bar P^{-\frac{1}{2}}(t_0\vert t_{-1})$ by the pseudo inverse of
$\bar P^{\frac{1}{2}}(t_0\vert t_{-1})$,
denoted by $\bar P^{\dag\frac{1}{2}}(t_0\vert t_{-1})$, and calculate it by singular value decomposition. In fact, according to \eqref{ensemble cov}, we have
\begin{equation}
 \bar P(t_0\vert t_{-1})=\dfrac{1}{q-1}\tilde X(t_0\vert t_{-1})\tilde X^T(t_0\vert t_{-1}).
\end{equation}
Then, by singular value decomposition,
\begin{equation}
\dfrac{1}{\sqrt{q-1}}\tilde X(t_0\vert t_{-1})=V_0S_0U_0^T,
\end{equation}
where $V_0\in \mathds R^{n\times n}$ and $U_0\in \mathds R^{q\times q}$ consist of the left and right singular vectors, $S_0\in \mathds R^{n\times q}$ is a rectangular diagonal matrix with singular values
$\{s_{0i}\vert s_{0i}\geqslant 0\}_{i=1}^q$ on the diagonal. Thus,
\begin{equation}
 \bar P(t_0\vert t_{-1})=V_0S_0U_0^TU_0S_0^TV_0^T=V_0S_0S_0^TV_0^T=V_0\hat S_0^2V_0^T,
\end{equation}
where $\hat S_0^2=S_0S_0^T \in \mathds R^{n\times n}$ is a block diagonal matrix with the rank $r_0$ and the diagonal $(s_{01}^2,\cdots,s_{0r_0}^2, 0_{1\times (n-r_0)})$.
Hence,
\begin{equation*}
 \bar P^{\dag\frac{1}{2}}(t_0\vert t_{-1})=V_0\hat S_0^\dag V_0^T,
\end{equation*}
where $\hat S_0^\dag$ is the pseudo inverse of $\hat S_0$ with the diagonal $(1/s_{01},\cdots,1/s_{0r_0}, 0_{1\times (n-r_0)})$. Then, the ensemble relative improvement covariance can be rewritten as
\begin{eqnarray*}
 &&\bar P^{\dag\frac{1}{2}}(t_0\vert t_{-1})(\bar P(t_0\vert t_{-1})-\bar P(t_0\vert t_N))\bar P^{\dag\frac{1}{2}}(t_0\vert t_{-1})\nonumber\\
 &=&\bar P^{\dag\frac{1}{2}}(t_0\vert t_{-1})\bar P(t_0\vert t_{-1})\bar P^{\dag\frac{1}{2}}(t_0\vert t_{-1})-\bar P^{\dag\frac{1}{2}}(t_0\vert t_{-1})\bar P(t_0\vert t_N)\bar P^{\dag\frac{1}{2}}(t_0\vert t_{-1})\\
 &=&V_0\hat S_0^\dag V_0^T(V_0\hat S_0^2V_0^T)V_0\hat S_0^\dag V_0^T-\bar P^{\dag\frac{1}{2}}(t_0\vert t_{-1})\bar P(t_0\vert t_N)\bar P^{\dag\frac{1}{2}}(t_0\vert t_{-1})\\
  &=&V_0I_{r_0}V_0^T-\bar P^{\dag\frac{1}{2}}(t_0\vert t_{-1})\bar P(t_0\vert t_N)\bar P^{\dag\frac{1}{2}}(t_0\vert t_{-1}),
\end{eqnarray*}
where $I_{r_0}$ is the diagonal matrix with the diagonal $(\mathds 1_{1\times r_0},0_{1\times (n-r_0)})$.

It is clear from \eqref{ensemble inverse cov} that $\bar P^{\dag\frac{1}{2}}(t_0\vert t_{-1})\bar P(t_0\vert t_N)\bar P^{\dag\frac{1}{2}}(t_0\vert t_{-1})$ is still nonnegative definite while $\bar P(t_0\vert t_{-1})$
is not with full rank, so if we use the same notation of standard Kalman filter and smoother to denote the ensemble relative improvement covariance, which means
\begin{equation*}
 \tilde P=\bar P^{\dag\frac{1}{2}}(t_0\vert t_{-1})(\bar P(t_0\vert t_{-1})-\bar P(t_0\vert t_N))\bar P^{\dag\frac{1}{2}}(t_0\vert t_{-1}),
\end{equation*}
then,
$
 0_{n\times n}\leqslant \tilde P<I_{r_0}.
$
Further, the \emph{ensemble relative improvement degree} is
\begin{equation}\label{ensemble relative degree}
\tilde p=\dfrac{\Vert \tilde P\Vert_1}{\Vert I_{r_0}\Vert_1}=\dfrac{\Vert \tilde P\Vert_1}{r_0}\in[0,1).
\end{equation}

As to the atmospheric transport model extended with emissions, for the distinction of the improvements for concentrations and emission rates, the \emph{ensemble relative ratios} are still
\begin{equation*}
\tilde p^c=\frac{\Vert \tilde P^c\Vert_1}{\Vert \tilde P \Vert_{1}},
\quad
\tilde p^e=\frac{\Vert \tilde P^e\Vert_1}{\Vert \tilde P \Vert_{1}}.
\end{equation*}

If we further consider the nonlinear dynamic model, we can renew the definition of the forecasting observation configurations as
\begin{equation*}
 y_k^{f}=\mathcal G (\hat x_k^{f}(t_0)),\ \ k=1,\cdots,q,
\end{equation*}
such that it can fully follow the nonlinear model, where $\mathcal G$ is a nonlinear operator.

Correspondingly, its ensemble mean and covariance are
\begin{equation*}
\bar y^f=\frac{1}{q}\sum_{k=1}^q y_k^{f}, \quad
\bar P_{yy}^f=\frac{1}{q-1}\sum_{k=1}^q(\hat y_k^f-\bar y^f)(\hat y_k^f-\bar y^f)^T.
\end{equation*}

Thus, for a nonlinear dynamic model, the extended ensemble Kalman filter to the theoretical approach in Section \ref{efficiency}, the only approximation is the limited size of the ensemble.

Hence, for the extended atmospheric transport model with emission rates, the analysis is similarto the analysis in Section \ref{extended model with emiss}.

\subsection{Example}
Consider a linear advection-diffusion model with periodic horizontal boundary condition and Neumann boundary condition in the vertical direction on the domain $ [0,14]\times [0,14]\times[0,4]$,
\begin{equation*}
\dfrac{\partial \delta c}{\partial t}=-v_x\dfrac{\partial \delta c}{\partial x}-v_y\dfrac{\partial \delta c}{\partial y}+\dfrac{\partial}{\partial z}(K(z)\dfrac{\partial \delta c}{\partial z})+\delta e-\delta d,
\end{equation*}
where $\delta c$, $\delta e$ and  $\delta d$ are the perturbations of the concentration, the emission rate and deposition rate of a species respectively.
$v_x$ and $v_y$ are constants and $K(z)$ is a differentiable
function of height $z$.

Assume $\triangle t=0.5$, the numerical solution is based on the symmetric operator splitting technique (\cite{Yanenko71}) with the following operator sequence
\begin{equation*}
\delta c(t+\triangle t)=T_xT_yD_zAD_zT_yT_x\delta c(t),
\end{equation*}
where $T_x$ and $T_y$ are transport operators in horizontal directions $(x,y)$, $D_z$ is the diffusion operator in  vertical direction $(z)$.
The parameters of emission and deposition rates are included in $A$. The Lax-Wendroff algorithm is chosen as the discretization method for horizontal advection with $\triangle x=\triangle y=1$. The vertical diffusion is discretized by Crank-Nicolson discretisation with the Thomas algorithm as solver.
The horizontal domain is $[0,14]\times [0,14]$ with the horizontal space discretization interval, while the vertical domain is   $[0,4]$ with $\triangle z=1$. So the number of the grid points $N_g=N_x\times N_y \times N_z=1125$, where $N_x=15$, $N_y=15$, $N_z=5$.

In addition, we choose $M_e(t,t_0)$, a continuous function in time $t$, to formulate the temporal background evolution profile shape of the emission rate as $$ e_b(t)=M_e(t,t_0)e_b(t_0),$$
where $e_b(t_0)$ is the initial value of emission rate.

With the same assumptions of $\triangle t$ and grid points in the 3D domain, the discrete dynamic model of emission rates is
\begin{equation*}
\delta e(t+\triangle t,i,j,l)=M_e(t+\triangle t,t)\delta e(t,i,j,l),
\end{equation*}
where  $\{(i,j,l), i,j\in \{0,\cdots, 14\}, l\in\{0,\cdots,4\}\}$ are the coordinates of grid points and $$M_e(t+\triangle t,t)=e_b(t+\triangle t)/ e_b(t).$$

For expository reasons the background assumption of $\delta d$ is denoted by $\delta d_b$, which is kept fixed.

According to the discretization of the phase space, we always assume there is only one fixed observation configuration in this example. It indicates that the observation operator mapping the state space to the observation space is a $1\times 2N_g$ time-invariant matrix.

Set 500 (the ensemble number $N$) samplings for the initial concentration and emission rate respectively by pseudo independent random numbers and make the states correlated by moving average technique.

\emph{Advection test:} For the advection test (Fig.~1 to Fig.~6), we assume the model with a weak diffusion process $(K(z)=0.5e^{-z^2})$ and
there is one single observation configuration of the concentration in the lowest layer at each time step,
denoted by 'Obs-cfg of conc' in figures. Besides, the emission source is assumed mainly from the location shown by the blue point in figures, named 'Emss-source'.

If we set the data assimilation window to $10\triangle t$ and the wind is from southwest, the left-side subplot in Fig.~1 shows the estimation of the concentration is probably improved at the field around the observation under the small assimilation window.
Meanwhile, though the right-side subplot in Fig.~1 shows  hardly improvement of the emission rate, we can see from the first line of Table 1 is feasible only for the concentration, for the simple reason that the single observation
configuration cannot detect the emission within the corresponding assimilation window.

If we consider the same case as Fig.~1, but now extending the data assimilation window to $35\triangle t$,
Fig.~2 shows the field where the concentration is potentially improved is enlarged since the states are more correlated with the extension of assimilation window and the estimation of
the emission surrounding the emission source is improved, compared to the Fig.~1. The quantitative balance between the concentration and the emission is shown by the relative improvement ratios
in the second line of Table 1.

If we further extend the data assimilation window to $48\triangle t$, it is clear to see from Fig.~3 and the third line of Table 1 that the states are more correlated such that more areas
can be analysed and improved by the single observation configuration. Meanwhile, the improvement of the emission is dominant with increasing time.

Fig.~4 to Fig.~6 show the relative improvements of the concentration and emission rate, when the model domain is under a northeasterly wind regime,  and assimilation windows of with the data assimilation windows $10\triangle t$, $35\triangle t$ and $48\triangle t$ respectively. It is easy to imagine that with  northeasterly winds, whatever the duration of the assimilation window is, the emission is not detectable and improveable by the single observation configuration. This hypothesis is successfully tested by our approach, the results of which are clearly visible in Fig.~4 to Fig.~6 and Table 2.

\emph{Emission signal test:} The purpose of emission signal test (Fig.~7 and Fig.~8) is to show the approach is also sensitive to the different background profile of the emission rate evolution.
Hence,  the only distinction between the situations in Fig.~7 and Fig.~8 is the background profile of the emission rate during the assimilation window
$48\triangle t$. Actually, Fig.~8 is the same case as Fig.~3. Thus, the result of the approach is clearly shown in Table 3 that the strong emission signal or the distinct variation of the emission rate during the data assimilation window is significant to the model to recognize the source of the changes of the concentration and improve the estimation of the states.

\emph{Diffusion test:} The diffusion test (Fig.~9 and Fig.~10) aims to test the approach via comparing the ensemble relative improvements of the concentration and the emission rate of the model with a weak diffusion process and a strong diffusion process. For the case in Fig.~9, all assumptions are same with the situation in Fig.~2 except that the single observation configuration is at the top layer instead.
The only difference of the assumptions between Fig.~9 and Fig.~10 is that $K(z)=0.5e^{-z^2}$ in Fig.~9 and $K(z)=0.5e^{-z^2}+1$ in Fig.~10.

Comparing Fig.~2 with Fig.~9, it is obvious that the different observation location influence on the distribution of the relative improvements of the concentration greatly. From Table 5, the total improvement value of the concentration in the lowest layer for Fig.~2 is shown to be larger than the one for Fig.~9. Besides, it can be seen in Table 4 that the observation configuration in the top layer cannot detect the emission with such weak diffusion under the assimilation window $35\triangle $.

If we compare Fig.~9 with Fig.~10, it is shown in Table 5 that both the total improvement value of the concentration in the lowest layer for Fig.~10 and the weight of the emission rate increase, which implies that the observation configuration is more efficient to detect the emission and improve the estimation of the state of the model with the strong diffusion in Fig.~10.

\begin{figure*}
\centering
\includegraphics[scale=0.9]{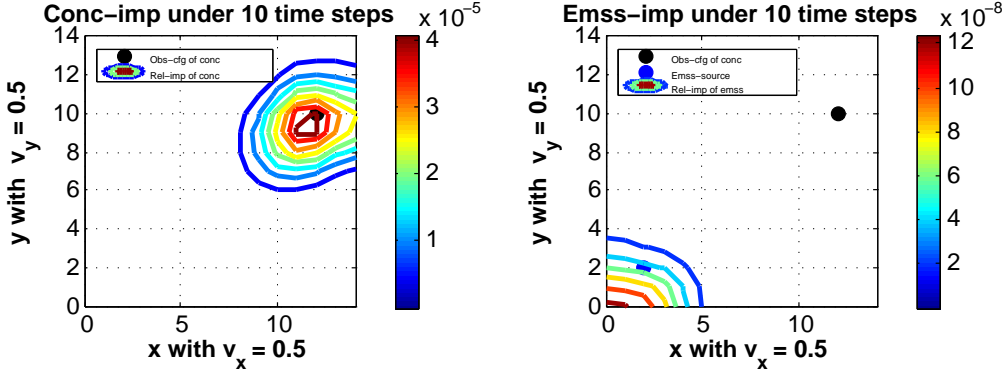}
\caption{Advection test with data assimilation (DA) window $10\triangle t$ and southwest wind. Isopleths of ensemble relative improvements of the concentration and emission rate are shown in the leftside and rightside figures respectively.}\qquad
\end{figure*}

\begin{figure*}
\centering
\includegraphics[scale=0.9]{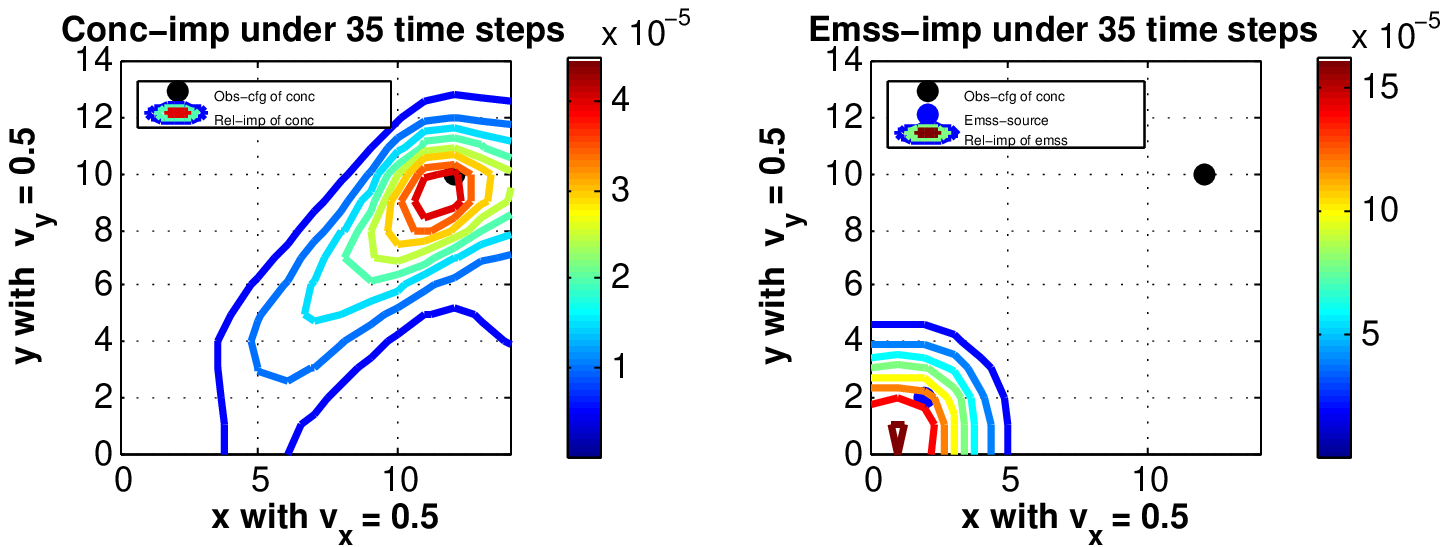}
\caption{Advection test with DA window $35\triangle t$ and southwest wind. Plotting conventions are as in Fig.~1.}
\end{figure*}

\begin{figure*}
\centering
\includegraphics[scale=0.9]{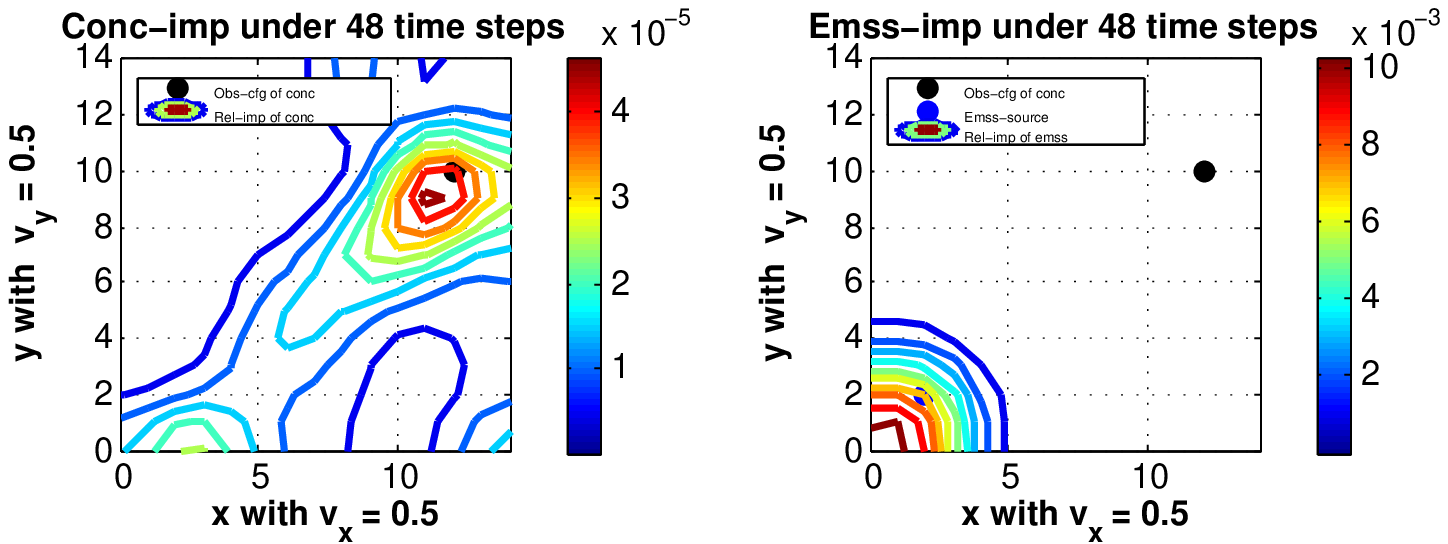}
\caption{Advection test with DA window $48\triangle t$ and southwest wind. Plotting conventions are as in Fig.~1.}
\end{figure*}

\begin{figure*}
\centering
\includegraphics[scale=0.9]{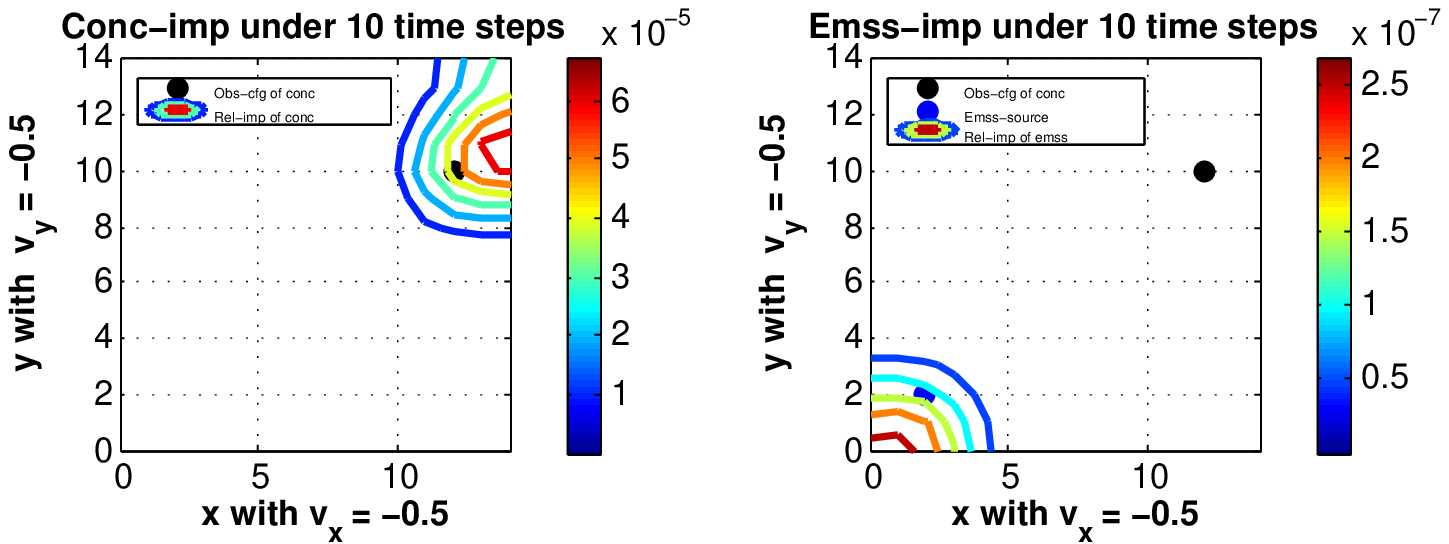}
\caption{Advection test with DA window $10\triangle t$ and northeast wind. Plotting conventions are as in Fig.~1.}
\end{figure*}

\begin{figure*}
\centering
\includegraphics[scale=0.9]{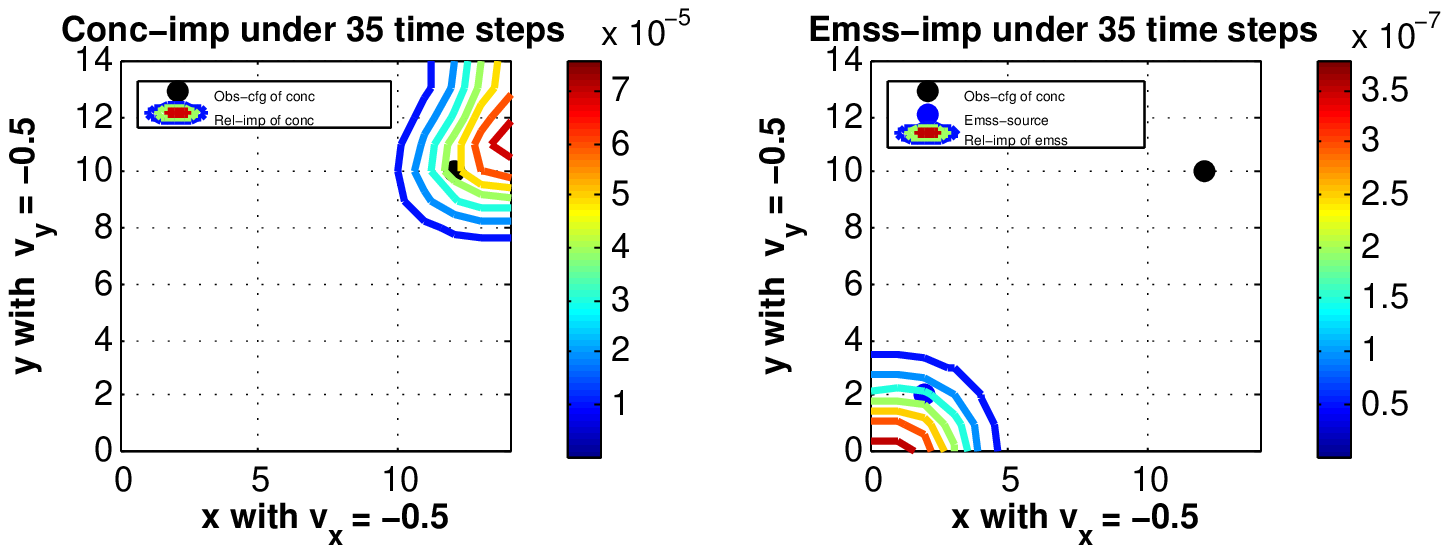}
\caption{Advection test with DA window $35\triangle t$ and northeast wind. Plotting conventions are as in Fig.~1.}
\end{figure*}

\begin{figure*}
\centering
\includegraphics[scale=0.9]{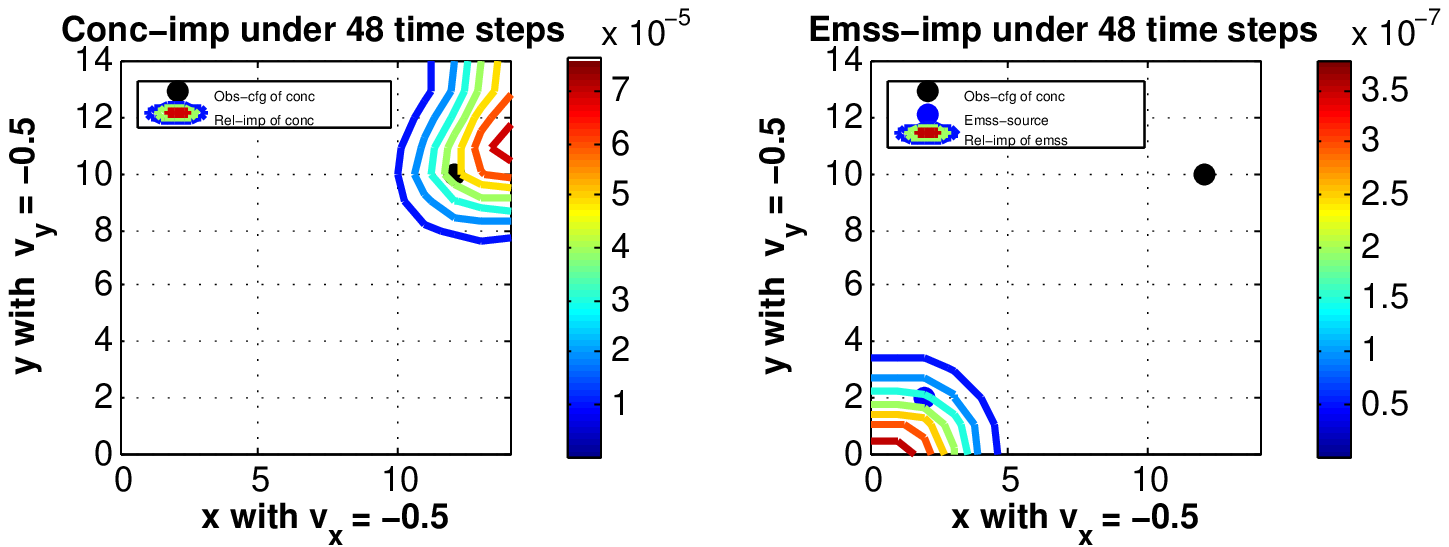}
\caption{Advection test with DA window $48\triangle t$ and northeast wind. Plotting conventions are as in Fig.~1.}
\end{figure*}

\begin{figure*}
\centering
\includegraphics[scale=0.7]{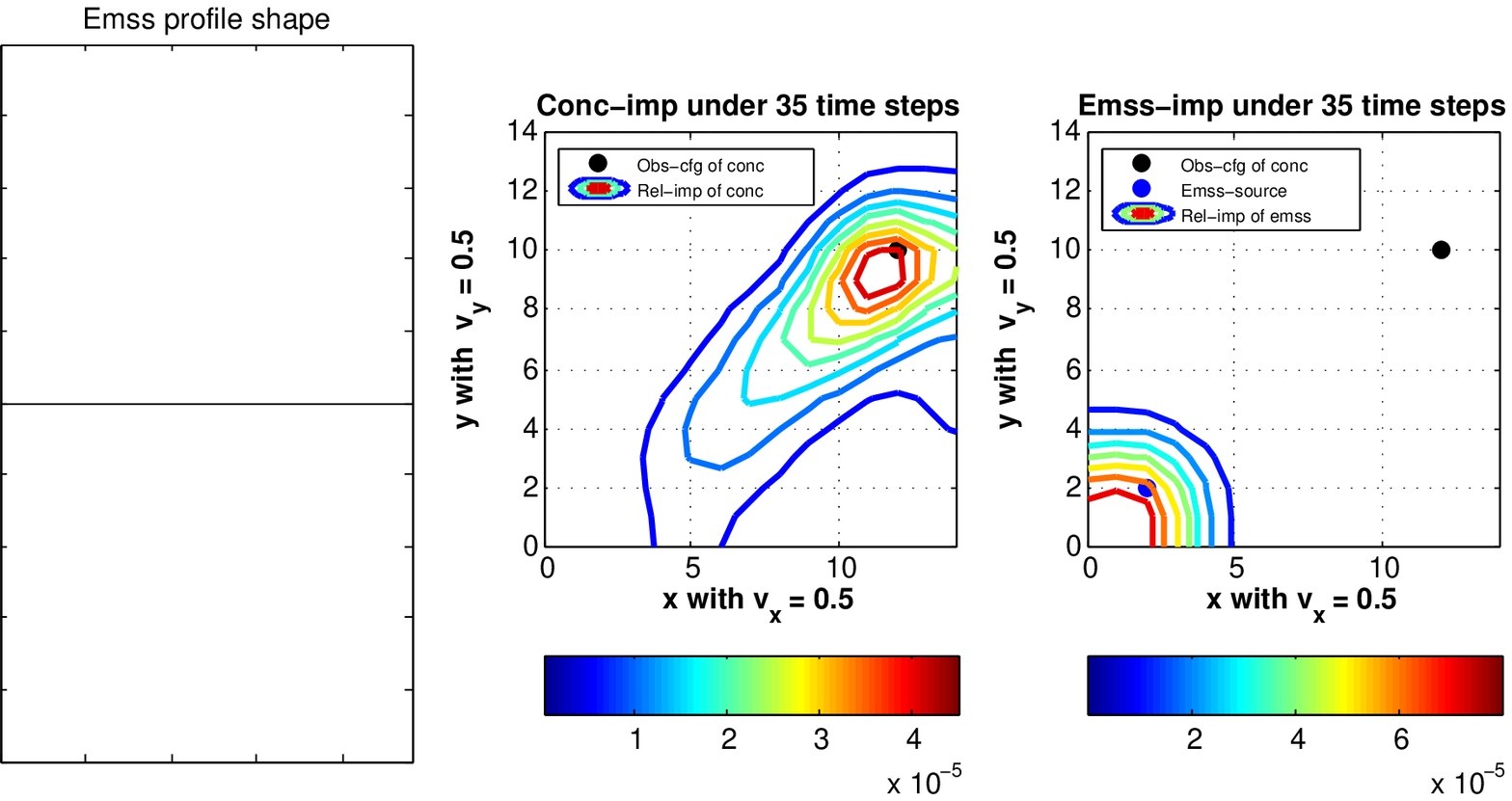}
\caption{Emission signal test (weak) with DA window $35\triangle t$ and southwest wind. Plotting conventions are as in Fig.~1.}
\end{figure*}

\begin{figure*}
\centering
\includegraphics[scale=0.7]{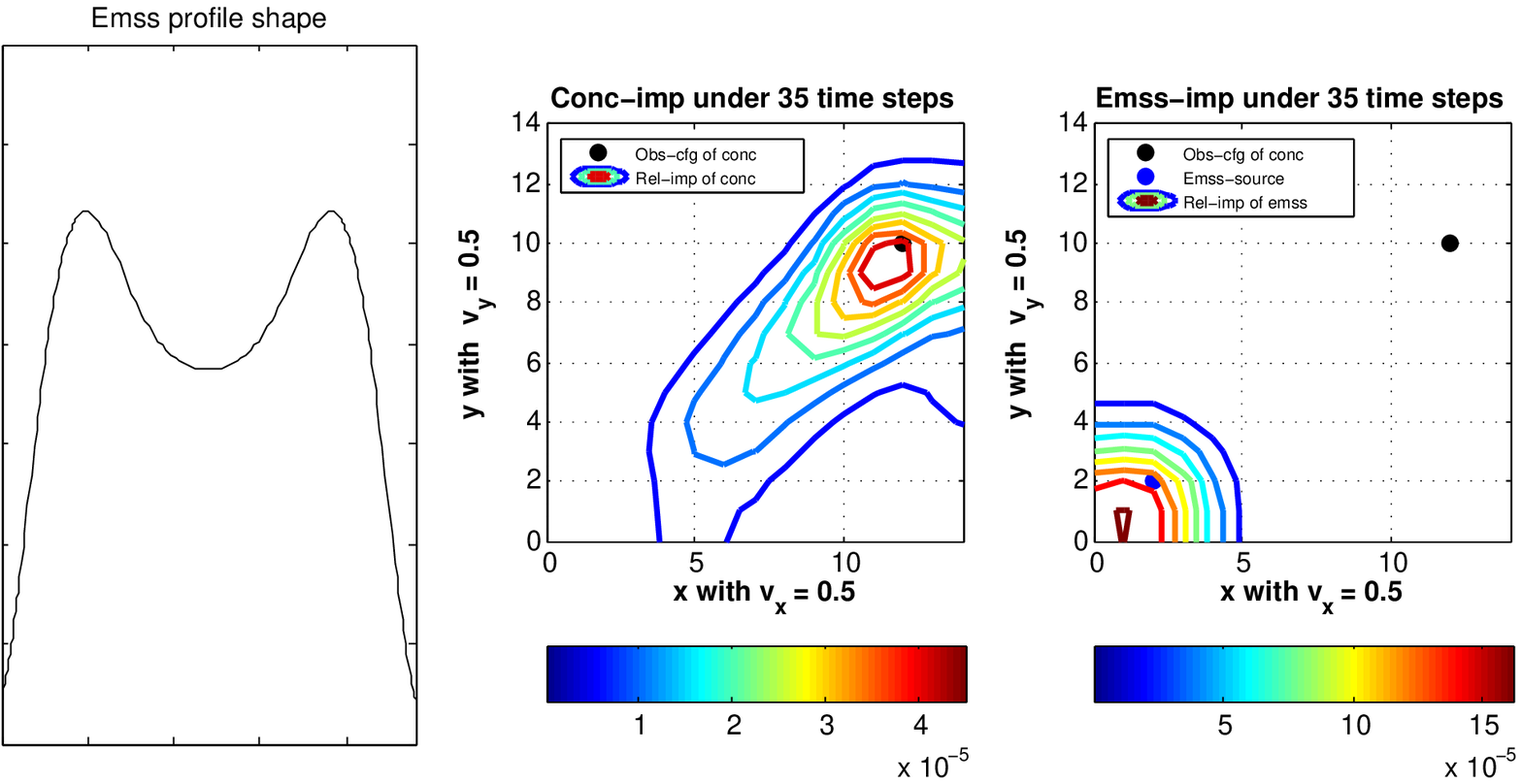}
\caption{Emission signal test (strong) with DA window $35\triangle t$ and southwest wind. Plotting conventions are as in Fig.~1.}
\end{figure*}

\begin{figure*}
\centering
\includegraphics[scale=0.8]{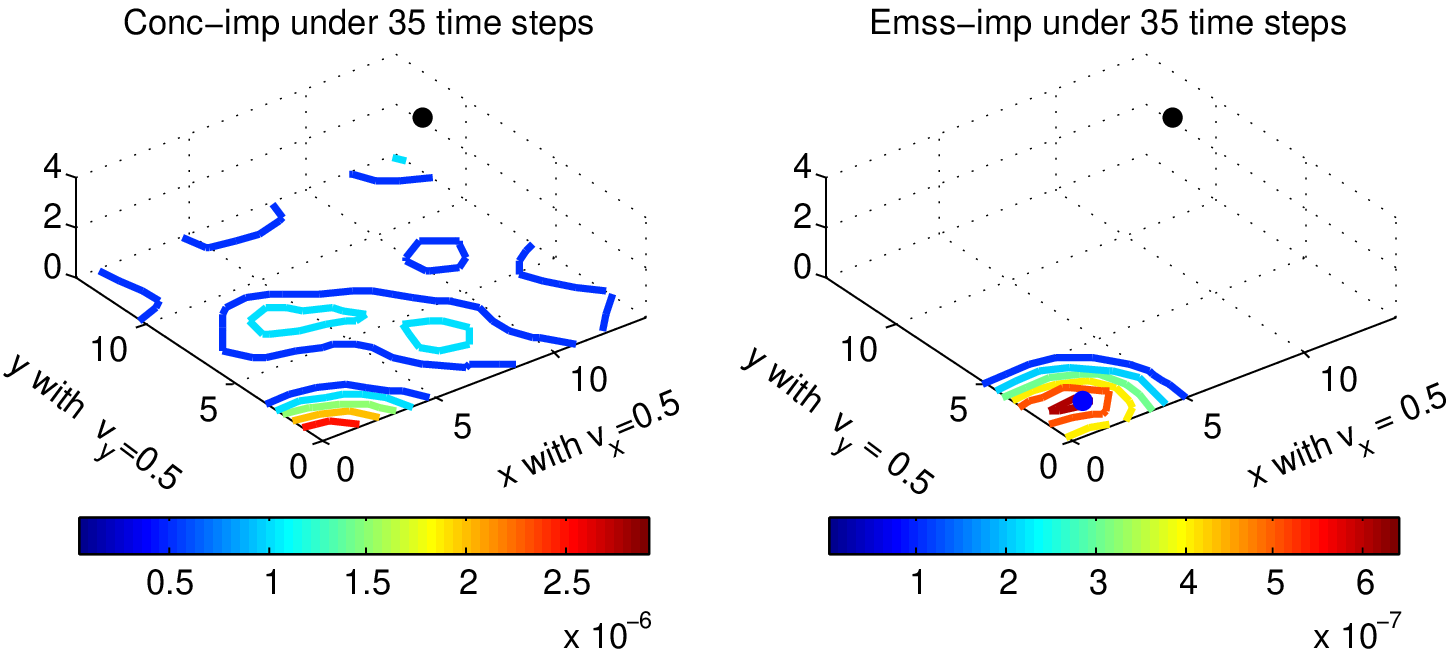}
\caption{Diffusion test (weak) with DA window $35\triangle t$ and southwest wind. Plotting conventions are as in Fig.~1.}
\end{figure*}

\begin{figure*}
\centering
\includegraphics[scale=0.8]{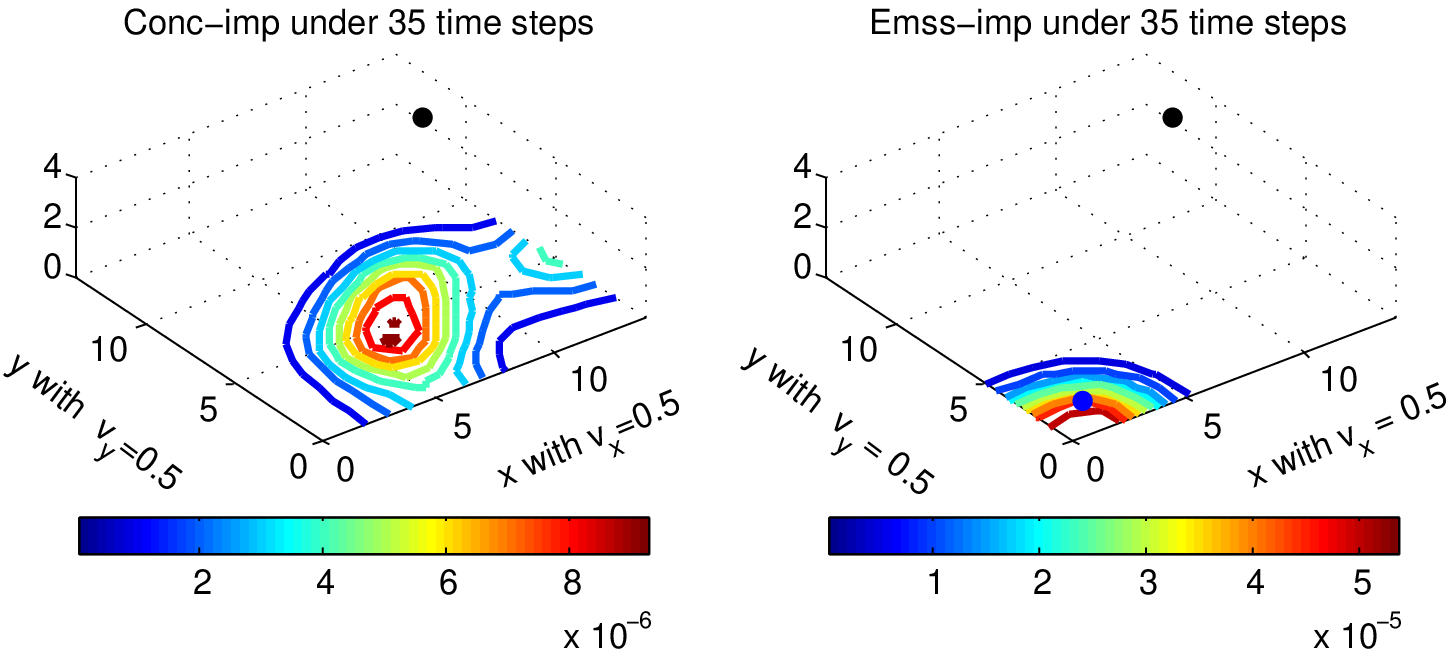}
\caption{Diffusion test (strong) with DA window $35\triangle t$ and southwest wind. Plotting conventions are as in Fig.~1.}
\end{figure*}
\begin{table}[ht]
 \footnotesize
   \centering
 \begin{minipage}[b]{.40\textwidth}
   \centering
 \begin{tabular}{ l | c   r }
  \hline\hline
  & $\tilde p^c$ & $\tilde p^e$ \\
\hline
  Fig.~1 & 0.9979 & 0.0021 \\
  Fig.~2 & 0.5107 & 0.4893 \\
  Fig.~3 &  0.0345 & 0.9655\\
  \hline\hline
\end{tabular}
  \caption{}
\end{minipage}\qquad
\begin{minipage}[b]{.40\textwidth}
  \centering
\begin{tabular}{ l | c   r }
  \hline\hline
  & $\tilde p^c$ & $\tilde p^e$ \\
\hline
  Fig.~4 & 0.9977 & 0.0023 \\
  Fig.~5 & 0.9974 & 0.0026 \\
  Fig.~6 &  0.9974 & 0.0026\\
  \hline\hline
\end{tabular}
  \caption{}
\end{minipage}
\end{table}

\begin{table}[ht]
 \footnotesize
   \centering
 \begin{minipage}[b]{.40\textwidth}
   \centering
\begin{tabular}{ l | c   r }
  \hline\hline
  & $\tilde p^c$ & $\tilde p^e$ \\
\hline
  Fig.~7 & 0.6811 & 0.3189 \\
  Fig.~8 & 0.5107 & 0.4893 \\
  \hline\hline
\end{tabular}
  \caption{}
\end{minipage}\qquad
\begin{minipage}[b]{.40\textwidth}
  \centering
\begin{tabular}{ l | c   r }
  \hline\hline
  & $\tilde p^c$ & $\tilde p^e$ \\
\hline
  Fig.~9 & 0.9977 & 0.0023 \\
  Fig.~10 & 0.7755 & 0.2245 \\
  \hline\hline
\end{tabular}
  \caption{}
\end{minipage}
\end{table}

\begin{table}[ht]
\centering
 \footnotesize
\begin{center}
\begin{tabular}{ l | c   r }
  \hline\hline
  & $\tilde P_{low}^c$ & $\tilde P_{low}^e$ \\
\hline
  Fig.~2 &  $2.8435\times 10^{-6}$ & $2.3530\times 10^{-6}$\\
  Fig.~9 & $2.3850\times 10^{-7}$& $2.1627\times 10^{-8}$ \\
  Fig.~10 & $7.8820\times 10^{-7}$& $1.5946\times 10^{-6}$ \\
  \hline\hline
\end{tabular}
\end{center}
\caption{$\tilde P_{low}^c$ and $\tilde P_{low}^e$ are respectively the total improvement values of the concentration and emission rates in the lowest layer}
\end{table}

\section{Sensitivity of observation networks to the initial state and emission rates}\label{sensitivity}

From the above discussion, we can determine the efficiency of the observation network by evaluating the improvement of estimation of initial state and emission rates separately,  before we run the data assimilation by Kalman filer and smoother.
However, it does not provide the information about the improved configurations of observations which can help improving the estimations.
In this section, independent of any concrete data assimilation method, we will introduce the singular vector approach to identify the sensitive directions of observations to the initial state and emission rates.

Consider the generalized discrete-time linear system:
\begin{equation*}
\delta x(t_{k+1})=M(t_{k+1},t_k)\delta x(t_k),
\end{equation*}
where $\delta  x(t_0)=x(t_0)-\hat x(t_0)$, $\hat x(t_0)$ is any estimate of $x(t_0)$.

Assume the observation mapping is accurate, which implies the data is the only source of observation errors, we have
\begin{equation*}
\delta y(t_k)=H(t_k)\delta x(t_k).
\end{equation*}

Define the magnitude of the perturbation of the initial state by the norm in the state space with respect to a positive definite matrix $W_0$
\begin{equation*}
\Vert \delta  x(t_0)\Vert_{W_0}^{2}=\langle \delta x(t_0), W_0\delta  x(t_0)\rangle.
\end{equation*}
Similarly, we define the magnitude of the related observations perturbation in the time interval $[t_0,\cdots, t_N]$ by the norm with respect to a sequence of positive definite matrix $\{W(t_k)\}_{k=1}^N$
$$
\Vert \delta y \Vert_{\lbrace W(t_k)\rbrace}^{2}=\sum_{k=0}^{N}\langle \delta y(t_k), W(t_k)\delta  y(t_k)\rangle,
$$
where $\delta y(t_k)=H(t_k)\delta x(t_k)$.

In order to find the direction of observation configuration which can minimize the perturbation of the initial states, the ratio
\begin{equation*}
\dfrac{{\Vert \delta x(t_0) \Vert_{W_0}^{2} }}{\Vert\delta y \Vert_{\lbrace W(t_k)\rbrace}^{2} }, \quad\delta y\neq 0
\end{equation*}
should be minimized. It is equivalent to maximize the ratio of the magnitude of observation perturbation and the initial perturbation
\begin{equation*}
\dfrac{\Vert \delta y \Vert_{\lbrace W(t_k)\rbrace}^{2}}{\Vert \delta x(t_0) \Vert_{W_0}^{2}},  \quad\delta x(t_0)\neq 0.
\end{equation*}

Thus, we define the measure the perturbation growth as
\begin{eqnarray}\label{dis ratio}
&&g^2= \dfrac{\Vert \delta y \Vert_{\lbrace W(t_k)\rbrace}^{2}}{\Vert \delta x(t_0) \Vert_{W_0}^{2}}\\
         &=& \sum_{k=0}^{N}\dfrac{\langle \delta y(t_k), W(t_k)\delta  y(t_k)\rangle}{\langle \delta  x(t_0), W_0\delta x(t_0)\rangle}\nonumber\\
         &=&\sum_{k=0}^{N}\dfrac{\langle H(t_k)\delta  x(t_k), W(t_k)H(t_k)\delta  x(t_k)\rangle}{\langle \delta x(t_0), W_0\delta x(t_0)\rangle}\nonumber\\
         &=&\sum_{k=0}^{N}\dfrac{\langle \delta  x(t_k),H(t_k)^{T}W(t_k)H(t_k)\delta  x(t_k)\rangle}{\langle \delta x(t_0), W_0\delta  x(t_0)\rangle}\nonumber\\
         &=& \sum_{k=0}^{N}\dfrac{\langle \delta  x(t_0), M(t_k,t_0)^{T}H(t_k)W(t_k)H(t_k)M(t_k,t_0)\delta  x(t_0)\rangle}{\langle \delta x(t_0), W_0\delta x(t_0)\rangle}\nonumber\\
         &=& \dfrac{\langle \delta  x(t_0),\sum_{k=0}^{N}M(t_k,t_0)^{T}H(t_k)^{T} W(t_k)H(t_k)M(t_k,t_0)\delta  x(t_0)\rangle}{\langle \delta x(t_0), W_0\delta x(t_0)\rangle}\nonumber.
\end{eqnarray}

According to Liao and Sandu (\cite{Liao06}), singular vectors refer to the directions of the error growth in a descend sequence with respect to the descent singular values. Hence, in order to search the maximal directions of
\begin{equation*}
g^2=\dfrac{\langle \delta x(t_0),\mathcal{G}^T\mathcal W\mathcal{G}\delta  x(t_0)\rangle}{\langle \delta x(t_0), W_0\delta x(t_0)\rangle},\ \  \delta x(t_0)\neq 0,
\end{equation*}
where $\mathcal W=\text{diag}(W(t_0),\cdots,W(t_N))$, $\mathcal G$ and $\mathcal R^{-1}$ have the same definitions with those in Section \ref{efficiency},
we need to find out the solutions of the singular value problem:
\begin{equation*}
W_0^{-\frac{1}{2}}\mathcal{G}^T\mathcal W\mathcal{G}W_0^{-\frac{1}{2}}v_k=s_{k}^2v_k,
\quad
\mathcal{G}W_0\mathcal{G}^Tu_k=s_{k}^2u_k,
\end{equation*}
where $s_1\geqslant s_2\geqslant\cdots\geqslant s_n\geqslant 0$, $\{v_k\}_{i=1}^n$ and $\{u_k\}_{i=1}^n$ are the corresponding orthogonal singular vectors.
Then, $
\max _{ \delta x(t_0)\neq 0}g^2=s_1^2.
$

Especially, if the perturbation norms are provided by the choice $W_0=P^{-1}(t_0\vert t_{-1})$ and $\mathcal W=\mathcal R^{-1}$ defined in Section \ref{efficiency},
\begin{equation*}
g^2=\dfrac{\langle \delta x(t_0),\mathcal{G}^T\mathcal R^{-1}\mathcal{G}\delta  x(t_0)\rangle}{\langle \delta x(t_0), P^{-1}(t_0\vert t_{-1})\delta x(t_0)\rangle},\ \  \delta x(t_0)\neq 0.
\end{equation*}
We need to search the directions of
\begin{equation}\label{sensitivity sv}
P^{\frac{1}{2}}(t_0\vert t_{-1})\mathcal{G}^T\mathcal R^{-1}\mathcal{G}P^{\frac{1}{2}}(t_0\vert t_{-1})v_k=s_{k}^2v_k;
\end{equation}
\begin{equation*}
\mathcal{G}P^{-1}(t_0\vert t_{-1})\mathcal{G}^Tu_k=s_{k}^2u_k, \ \ k=1,\cdots, n.
\end{equation*}

Associated with \eqref{improve measure2}, it is easy to find that the singular vector $v_k$ in \eqref{sensitivity sv},
which is the direction of $\text{k}^{\text{th}}$-fast growth of the perturbation of observations evolved from the initial perturbation,
is also the $\text{k}^{\text{th}}$ direction which maximize the improvement of estimation by Kalman filter and smoother \eqref{improve measure2},
though the exact value of the eigenvalue of \eqref{improve measure2} related to $v_k$  is $\frac{s_k^2}{1+s_k^2}$ rather than the eigenvalue $s_k^2$ of \eqref{sensitivity sv}.
Meanwhile, we can find that the leading singular value $s_1$ is related to the operator norm of $\tilde P$ as
\begin{equation*}
\Vert\tilde P\Vert=\max_{\Vert x\Vert=1}\Vert \tilde Px\Vert=\frac{s_1^2}{1+s_1^2}.
\end{equation*}
In addition, the similar analysis for the continuous-time system is presented in the appendix B.

\section{Discussion}
In the present work, approaches for determining the efficiency and sensitivity of observation configurations for the initial state and emission rates
are established.
Actually, to deal with the specific questions in atmospheric chemistry, some special operators are usually applied.
For example, in order to consider the efficiency and sensitivity of observations in some certain locations, the local projection operator introduced by Buizza et al. (\cite{Buizza99}) can be applied
into approaches in Section \ref{ensemble efficiency} and Section \ref{sensitivity}.

Let $L$ be the $0-1$ diagonal matrix defined as
\begin{equation*}
L_{ii}=\lbrace
\begin{array}{cc}
1,& \ l_i \in L_a ,\\
0,& \text{otherwise.}
\end{array}
\end{equation*}
where $L_a$ is a fixed area and $l_i$ is the coordinate of $i^{\text{th}}$ grid point.

To test the efficiency and sensitivity of observation configurations in a special area, by rearranging the observations $y$ according to the locations, $\mathcal G$ in \eqref{def G} should be defined as
\begin{equation}
\mathcal{G}=\left(\begin{array}{cc}
      LH(t_0)M(t_0,t_0)\\
      LH(t_1)M(t_1,t_0)\\
      \vdots\\
      LH(t_N)M(t_N,t_0)
      \end{array}
\right).
\end{equation}
If $LH(\cdot)$ is considered as the observation mapping,  approaches in Section \ref{efficiency} and \ref{sensitivity} can be applied.

In addition, if there is a multiplication of emission rates in the following model
\begin{equation*}
\frac{d\delta c}{dt}=\mathbf{A}\delta c+ B(t)\delta e(t),
\end{equation*}
then all approaches can also be applied into the extended model
\begin{equation*}
\left(\begin{array}{c}
      \delta c(t)\\
     \delta e(t)
      \end{array}
\right)=
\left(\begin{array}{cc}
      M(t,t_0) & \int_{t_0}^t M(t, s)B(s)M_e(s,t_0)ds\\
      0     & M_e(t,t_0)
      \end{array}
\right)
\left(\begin{array}{c}
       \delta c(t_0)\\
     \delta e(t_0)
      \end{array}
\right).
\end{equation*}

\appendix\section{The efficiency of observation networks for continuous-time systems}
Consider the abstract continuous-time system
\begin{align*}
&x(t)=M(t,t_0)x(t_0)+\varepsilon(t),\\
&y(t)=H(t)x(t)+\nu(t),
\end{align*}
where $x\in \mathds R^n$  is the state variable, $y\in \mathds R^{m}$ is  observation vector at time $t$,  the model error $\varepsilon(t)$ and the observation error
$\nu(t)$, $ t\in [t_0, t_N]$ follow  Gaussian distribution with zero mean, while
$Q(t)$ and $R(t)$ are their covariance matrices respectively.

As in Section \ref{efficiency}, we ignore the model error. It is well known that for the continuous Kalman filter, the covariance $P(t)$ of the optimal estimation of the state at time $t$ satisfies the integral Riccati equation
\begin{equation*}
P(t\vert t)=M_K(t,t_0)P(t_0\vert t_{-1})M_K^T(t,t_0)+\int_{t_0}^tM_K(t,s)K(s)R(s)K^T(s)M_K^T(t,s)ds,
\end{equation*}
where $K(t)=P(t\vert t)H(t)R^{-1}(t)$ and
$M_K(t,t_0)=M(t,t_0)-\int_{t_0}^tM(t,s)K(s)H(s)M_K(s,t_0)ds$.

On one hand,
\begin{eqnarray*}
&&M_K(t,t_0)P(t_0\vert t_{-1})M_K^T(t,t_0)\\
&=&M(t,t_0)P(t_0\vert t_{-1})M_K^T(t,t_0)-\int_{t_0}^tM(t,s)K(s)H(s)M_K(s,t_0)P(t_0\vert t_{-1})M_K^T(t,t_0)ds\\
&=&M(t,t_0)P(t_0\vert t_{-1})M_K^T(t,t_0)-\int_{t_0}^tM(t,s)K(s)H(s)P(s\vert s)M_K^T(t,s)ds\\
&&+\int_{t_0}^t\int_{t_0}^sM(t,s)K(s)H(s)M_K(s,\eta)K(\eta)R(\eta)K^T(\eta)M_K^T(t,\eta)d\eta ds. \\
\end{eqnarray*}
On the other hand,
\begin{eqnarray*}
&&\int_{t_0}^tM_K(t,s)K(s)R(s)K^T(s)M_K^T(t,s)ds\\
&=&\int_{t_0}^t[M(t,s)-\int_{s}^t M(t,\eta)K(\eta)H(\eta)M_K(\eta,s)d\eta]K(s)R(s)K^T(s)M_K^T(t,s)ds\\
&=&\int_{t_0}^tM(t,s)K(s)R(s)K^T(s)M_K^T(t,s)ds\\
&&-\int_{t_0}^t\int_{0}^\eta M(t,\eta)K(\eta)H(\eta)M_K(\eta,s)K(s)R(s)K^T(s)M_K^T(t,s)dsd\eta\\
&=&\int_{t_0}^tM(t,s)K(s)R(s)K^T(s)M_K^T(t,s)ds\\
&&-\int_{t_0}^t\int_{0}^s M(t,s)K(s)H(s)M_K(s,\eta)K(\eta)R(\eta)K^T(\eta)M_K^T(t,\eta)d\eta ds.
\end{eqnarray*}
Therefore,
$P(t\vert t)=M(t,t_0)P(t_0\vert t_{-1})M_K^T(t,t_0)$.

Since
\begin{eqnarray*}
&&M^{-1}(t,t_0)=M_K^{-1}(t,t_0)M_K(t,t_0)M^{-1}(t,t_0)\\
&=&M_K^{-1}(t,t_0)[M(t,t_0)-\int_{t_0}^tM_K(t,s)K(s)H(s)M(s,t_0)ds]M^{-1}(t,t_0)\\
&=&M_K^{-1}(t,t_0)-\int_{t_0}^tM_K^{-1}(s,t_0)L(s)H(s)M(t,s)ds,
\end{eqnarray*}
we obtain $M_K^{-1}(t,t_0)=M^{-1}(t,t_0)+\int_{t_0}^tM_K^{-1}(s,t_0)K(s)H(s)M^{-1}(t,s)ds$.

Define $\hat x(t_0\vert t)=E[x(t_0)\vert y(s),s\in[t_0,t]]$ and denote its covariance matrix as
$$P(t_0\vert t)=E[(x(t_0)-\hat x(t_0\vert t))(x(t_0)-\hat x(t_0\vert t))^T],$$
Hence,
\begin{eqnarray*}
&&P^{-1}(t_0\vert t)\\\nonumber
&=&[M^{-1}(t,t_0)P(t\vert t)M^{-T}(t,t_0)]^{-1}\\\nonumber
&=&[P(t_0\vert t_{-1})M_K^{T}(t,t_0)M^{-T}(t,t_0)]^{-1}\\\nonumber
&=&M^{T}(t,t_0)[M^{-T}(t,t_0)+\int_{t_0}^tM^{-T}(t,s)H^T(s)K^T(s)M_K^{-T}(s,t_0)ds]P(t_0\vert t_{-1})^{-1}\\\nonumber
&=& P^{-1}(t_0\vert t_{-1})+\int_{t_0}^tM^T(s,t_0) H^T(s)R^{-1}(s)H(s)M(s,t_0)ds.
\end{eqnarray*}

Let $t=t_N$ and define the observability mapping $\mathcal{G}: \mathds R^n\rightarrow \mathcal L^{2}([t_0, t_N]; \mathds R^m)$ as
$$
\mathcal{G}f:=H(\cdot)M(\cdot,t_0)f, \quad f\in\mathds R^n,
$$
its adjoint operator $\mathcal{G}^{*}$ is
$$
\mathcal{G}^{*}f=-\int_{t_N}^{t_0}M^{T}(s,t_0)H^{T}(s)f(s)ds, \quad f\in  L^{2}([t_0, t_N]; \mathds R^m).
$$
Further, we define $\mathcal R^{-1}:\mathcal L^{2}([t_0, t_N]; \mathds R^m)\rightarrow\mathcal L^{2}([t_0, t_N]; \mathds R^m)$,
$$
\mathcal{R}^{-1}f:=R^{-1}(\cdot)f(\cdot),\quad f\in \mathcal L^{2}([t_0, t_N]; \mathds R^m).
$$
Thus,
\begin{equation}\label{continuous cov}
P^{-1}(t_0\vert t_N)=P^{-1}(t_0\vert t_{-1})+\mathcal{G}^{*}\mathcal R^{-1}\mathcal{G},
\end{equation}
where $\mathcal{G}^{*}\mathcal R^{-1}\mathcal{G}$ is the observability Gramian of continuous-time systems.

Obviously, \eqref{continuous cov} has the same pattern as \eqref{inverse smoother}, so
\begin{eqnarray}\label{con improve measure}
&&P^{-\frac{1}{2}}(t_0\vert t_{-1})(P(t_0\vert t_{-1})-P(t_0\vert t_N))P^{-\frac{1}{2}}(t_0\vert t_{-1})\nonumber\\
&=&I-(I+P^{\frac{1}{2}}(t_0\vert t_{-1})\mathcal{G}^{T}\mathcal R^{-1}\mathcal{G}P^{\frac{1}{2}}(t_0\vert t_{-1}))^{-1}\nonumber\\
&=& V(I-(I+S^2)^{-1}) V^T,
\end{eqnarray}
where $VS^2V^T$ is the singular value decomposition of $P(t_0\vert t_{-1})^{\frac{1}{2}}\mathcal{G}^{T}\mathcal R^{-1}\mathcal{G}P(t_0\vert t_{-1})^{\frac{1}{2}}$.

Then, following the same steps as in Section \ref{efficiency}, we can obtain the efficiency of observation configurations for continuous-time systems.

\section{The sensitivity of observation networks for continuous-time systems}

Consider the generalized continuous-time linear system:
\begin{equation*}
\delta x(t)=M(t,t_0)\delta x(t_0),
\end{equation*}
with the corresponding forecast perturbation of observations evolving from $\delta x(t_0)$
\begin{equation*}
\delta y(t)=H(t)\delta x(t).
\end{equation*}

To be brief, let $W_0=P^{-1}(t_0\vert t_{-1})$ and $\mathcal W(t)=R^{-1}(t)$ (see Appendix A), and define the magnitude of the perturbation of the initial state and observations respectively by
\begin{equation*}
\Vert \delta  x(t_0)\Vert_{P^{-1}(t_0\vert t_{-1})}^{2}=\langle \delta x(t_0), P^{-1}(t_0\vert t_{-1})\delta  x(t_0)\rangle,
\end{equation*}
$$
\Vert \delta y \Vert_{\lbrace R^{-1}(t)\rbrace}^{2}=\int_{t_0}^{t_N}\langle \delta y(t), R^{-1}(t)\delta  y(t)\rangle dt.
$$

Thus, the perturbation growth for continuous-time system can be measured by
\begin{eqnarray*}\label{con ratio}
&&g^2= \dfrac{\Vert \delta y \Vert_{\lbrace R^{-1}(t))\rbrace}^{2}}{\Vert \delta x(t_0) \Vert_{P^{-1}(t_0\vert t_{-1})}^{2}}\nonumber\\
         &=&\dfrac{\int_{t_0}^{t_N}\langle H(t)\delta  x(t), R^{-1}(t)H(t)\delta  x(t)\rangle dt}{\langle \delta x(t_0), P^{-1}(t_0\vert t_{-1})\delta x(t_0)\rangle}\nonumber\\
         &=& \dfrac{\langle \delta  x(t_0),\int_{t_0}^{t_N}M(t,t_0)^{T}H(t)^{T}R^{-1}(t)H(t)M(t,t_0)\delta  x(t_0) dt\rangle}{\langle \delta x(t_0), P^{-1}(t_0\vert t_{-1})\delta x(t_0)\rangle}\nonumber\\
         &=&\dfrac{\langle \delta x(t_0),\mathcal{G}^*\mathcal R^{-1}\mathcal{G}\delta  x(t_0)\rangle}{\langle \delta x(t_0), P^{-1}(t_0\vert t_{-1})\delta x(t_0)\rangle},\ \ \ \delta x(t_0)\neq 0,
\end{eqnarray*}
where  $\mathcal G$ and $\mathcal R^{-1}$ are defined in Appendix A.

To find the directions maximizing the ratio, we need to find the solutions of the singular value problem:
\begin{align}\label{sensitivity sv2}
&P^{\frac{1}{2}}(t_0\vert t_{-1})\mathcal{G}^T\mathcal R^{-1}\mathcal{G}P^{\frac{1}{2}}(t_0\vert t_{-1})v_k=s_{k}^2v_k,\\
&\mathcal{G}P^{-1}(t_0\vert t_{-1})\mathcal{G}^Tu_k=s_{k}^2u_k.
\end{align}
where $s_1\geqslant s_2\geqslant\cdots\geqslant s_n\geqslant 0$, $\{v_k\}_{i=1}^n$ and $\{u_k\}_{i=1}^n$ are orthogonal singular vectors.

Compared \eqref{sensitivity sv2} with \eqref{con improve measure} in Appendix A, similar analysis and conclusions as Section \ref{sensitivity} can be extended to continuous-time systems.

\end{document}